\documentclass{aa}
\usepackage{txfonts}
\usepackage{graphicx}
\usepackage{natbib}
\usepackage{longtable,lscape}
\bibpunct{(}{)}{;}{a}{}{,}

\begin{document}

\title{A photometric and spectroscopic study of the new dwarf spheroidal galaxy in Hercules
 \thanks{Based on observations made with the INT telescope operated
    on the island of La Palma by the Isaac Newton Group in the Spanish
    Observatorio del Roque de los Muchachos of the Instituto de
    Astrofisica de Canarias.  }\fnmsep\thanks{Based on spectra
    obtained with the VLT-U2 telescope ESO Proposal number
    079.B-0447(A).}}

\subtitle{Metallicity, velocities and a clean list of RGB members}

\author{ D.\,Ad\'{e}n\inst{1} \and S.\,Feltzing\inst{1} \and A.\,Koch\inst{2} \and
  M.I.\,Wilkinson\inst{2} \and E.K.\,Grebel\inst{3} \and
  I.\,Lundstr\"om\inst{1}\and G.F.\,Gilmore\inst{4} \and  D.B.\,Zucker\inst{5,6} \and V.\,Belokurov\inst{4} \and N.W.\,Evans\inst{4} \and
  D.\,Faria\inst{4} }

\authorrunning{Ad\'en et al.}

\offprints{D. Ad\'{e}n  daniel.aden@astro.lu.se }
 
\institute{Lund Observatory, Box 43, SE-22100 Lund, Sweden \and
  Deptartment of Physics and Astronomy, University of Leicester,
  University Road, Leicester LE1 7RH, UK \and
Astronomisches Rechen-Institut, Zentrum f\"ur Astronomie der 
Universit\"at Heidelberg, M\"onchhofstr. 12-14, 69120 Heidelberg, Germany\and
  Institute of Astronomy, Madingley Road, Cambridge, CB3 0HA, UK\and Department of Physics, Macquarie University, North Ryde, NSW 2109, Australia
  \and Anglo-Australian Observatory, PO Box 296, Epping, NSW 1710, Australia}

\date{Received 18 June 2009 / Accepted 13 August 2009}

\abstract {dSph galaxies are the least luminous, least massive galaxies known. Recently,
the number of observed galaxies in this class has greatly increased thanks to large surveys. 
Determining their properties, such as mass, luminosity and metallicity,
provides key information in our understanding of galaxy formation and
evolution. } 
{
Our aim is to provide as clean and as complete a sample as possible
of red giant branch stars that are members of the Hercules dSph galaxy.
With this sample we explore the velocity dispersion and the metallicity of the system.
} 
{
Str\"omgren photometry and multi-fibre spectroscopy are combined to 
provide information about the evolutionary state of the stars (via the Str\"omgren 
c$_{\rm 1}$ index) and their radial velocities. Based on this information
we have selected a clean sample of red giant branch stars, and show
that foreground contamination by Milky Way dwarf stars can greatly
distort the results.}
{Our final sample consists of 28 red giant branch stars in the Hercules
dSph galaxy. Based on these stars we find a mean photometric
metallicity of $-2.35 \pm 0.31$ dex which is consistent with previous studies. We find evidence for an abundance spread.
Using those stars for which we have determined radial velocities we find a systemic velocity of $45.20 \pm 1.09 {\rm \, km\, s^{-1}}$
with a dispersion of 3.72 km\,s$^{-1}$, this is lower than values found in the literature. 
Furthermore we identify the horizontal branch and estimate the mean magnitude of the horizontal branch of the Hercules dSph galaxy to be $V_0=21.17 \pm
0.05$, which corresponds to a distance of $147^{+8}_{-7}$ kpc.} 
{When studying sparsely populated and/or heavily foreground contaminated
dSph galaxies it is necessary to include knowledge about the evolutionary
stage of the stars. This can be done in several ways. Here we have 
explored the power of the $c_{{\rm 1}}$ index in Str\"omgren photometry. This 
index is able to clearly identify red giant branch stars redder than
the horizontal branch, enabling a separation of red giant branch dSph stars and foreground dwarf stars. Additionally, this index is also capable of 
correctly identifying both red and blue horizontal branch stars. 
We have shown that a proper cleaning of the sample results in a smaller value for the velocity dispersion of the system. This has implications for galaxy properties
derived from such velocity dispersions.}

\maketitle

\section{Introduction}  \label{a}

In the past few years the number of known Milky Way satellites
has increased considerably. Our Galaxy has gained at least ten newly
recognized companions, and additional ones await confirmation
\citep[e.g.][]{2006ApJ...650L..41Z,2006ApJ...643L.103Z,2007ApJ...662L..83W,2006ApJ...647L.111B,2007ApJ...654..897B,2008ApJ...686L..83B,2009MNRAS.397.1748B}.
These recently discovered satellites resemble the previously known
dwarf spheroidal galaxies in many of their characteristics \citep[see][for a
detailed discussion of the properties of the classical dSph galaxies]{2003AJ....125.1926G}, but
most of the new satellites are several magnitudes fainter than any dwarf galaxy known
before. Hence these objects are now often referred to as ``ultra-faint dwarf
spheroidal galaxies''. Since most of the new discoveries were made using
deep CCD sky surveys such as the Sloan Digital Sky Survey, which cover
primarily the northern hemisphere, it seems highly likely that additional
objects will be added once the southern sky is scanned in a similar manner
\citep[e.g.][]{2007PASA...24....1K,2009AJ....137..450W}.

One of the new discoveries is the Hercules dwarf spheroidal (dSph) galaxy
\citep{2007ApJ...654..897B}.  Hercules lies at a distance of $132\pm12$ kpc from
us, has an absolute $V$-band magnitude of about $-6.6\pm0.3$, a $V$-band surface
brightness of only $27.2\pm0.6$ mag arcsec$^{-2}$, and appears highly elongated
 \citep{2007ApJ...668L..43C,2008ApJ...684.1075M}.  Its stellar mass is
estimated
to be in the range of several $\times 10^4$ M$_{\odot}$.  The mass of the
Hercules dSph galaxy, as
inferred from line-of-sight radial velocity measurements, is
of the order of $10^7$  M$_{\odot}$ within the central 300 pc
\citep{2008Natur.454.1096S}.  This value is in good agreement with the
seemingly ubiquitous,
common mass scale of the other Galactic satellites \citep[e.g.][and
references
therein]{2007ApJ...663..948G,2007ApJ...667L..53W,2008Natur.454.1096S}.

As would be expected from its low luminosity and low surface
brightness, Hercules is a metal-poor dSph galaxy.  {From} measurements
of the Ca {\sc ii} IR triplet lines in red giant stars, \citet{2007ApJ...670..313S} inferred a mean metallicity of
[Fe/H] = $-2.27\pm0.07$ on the metallicity scale of
\citet{1997A&AS..121...95C}. Using spectrum synthesis of Fe\,{\sc i} lines,
\citet{2008ApJ...685L..43K} derived a mean metallicity of
$-2.58\pm0.51$ dex. Both studies found a wide range of metallicities
among the red giant stars in Hercules, confirming the trend known from
brighter dSph galaxies that often exhibit spreads of 1 dex and more
\citep[e.g.][]{2001ApJ...548..592S,2006AJ....131..895K,2007AJ....133..270K,2007ApJ...657..241K}.

A detailed, high-resolution abundance analysis of two red giants in
the Hercules dSph galaxy revealed that the enrichment in heavy
elements proceeded inhomogeneously and that core-collapse supernovae 
were the primary contributors to the enrichment of Hercules \citep{2008ApJ...688L..13K}. 
Evidence for such chemical
inhomogeneities on small scales has also been found in the more
luminous dSph galaxies \citep{2008AJ....135.1580K} and in more massive dIrr galaxies \citep{2005AJ....130.1558K}.

In our current study we explore the potential of Str\"omgren
photometry for the study of the stellar content of ultra-faint dSph galaxies, such as the one in
Hercules.  In contrast to the usual broadband photometry, Str\"omgren
photometry offers several advantages. It provides us with
gravity-sensitive multi-colour indices useful for distinguishing different
evolutionary stages including giant-dwarf discrimination. This is a
very valuable option when trying to eliminate foreground dwarfs from a
giant candidate sample to be used for subsequent spectroscopy,
particularly when dealing with sparse, extended, ultra-faint dSph
galaxies that tend to suffer from substantial Galactic foreground
contamination.  Moreover, Str\"omgren indices offer the possibility to
estimate metallicities for red giant stars.  This method has been
considerably refined since the early calibration attempts by
\citet{1992A&A...253..359G}.  As compared to other photometric
estimates, it has the added advantage of providing age-independent
metallicity estimates \citep[e.g.][]{2007A&A...465..357F}.  In terms
of telescope time, intermediate-band photometry is a lot cheaper than
spectroscopic surveys of faint giants.   \\

We have obtained both Str\"omgren photometry as well as spectroscopic
observations of the Ca {\sc ii} IR triplet lines for stars in the field of the
Hercules dSph galaxy. These observations enable a full analysis both
of evolutionary stage as well as radial velocities for these stars. It
turns out that knowledge about the evolutionary stage of the stars is
crucial for the construction of a clean sample of red giant branch (RGB) stars in the
Hercules dSph galaxy.

The paper is organized as follows: In Sect. \ref{b} we describe the
observations and data reductions for both the photometric and the
spectroscopic observations. In Sect. \ref{result} we present the
colour-magnitude diagram as well as the measured radial velocities. In
Sect. \ref{findrgb} we show how the gravity sensitive Str\"omgren
$c_1$ index can be used to disentangle the Hercules dSph galaxy
members from the foreground contamination, and Sect. \ref{vrmemb}
deals with membership determination based on radial velocities. 
Section \ref{fs} summarizes how we define a Hercules member star.
Section \ref{vrcomp} provides a comparison with previous velocity determinations.
In Sect. \ref{metal} we derive metallicities for the stars identified as
members of the Hercules dSph galaxy using the Str\"omgren $m_1$ index
as well as from the measurements of the Ca {\sc ii} IR triplet lines, and
compare the results with previous studies of the metallicity of the
Hercules dSph galaxy. In
Sect. \ref{disc} our results are discussed and they are summarized in Sect. \ref{k}.
 
	\section{Observations and data reductions}  \label{b}
In this section we detail the observations as well as the data reduction for both photometric and spectroscopic observations.
	\subsection{Photometry}  \label{cc}
	
\begin{figure}    
\resizebox{\hsize}{!}{\includegraphics{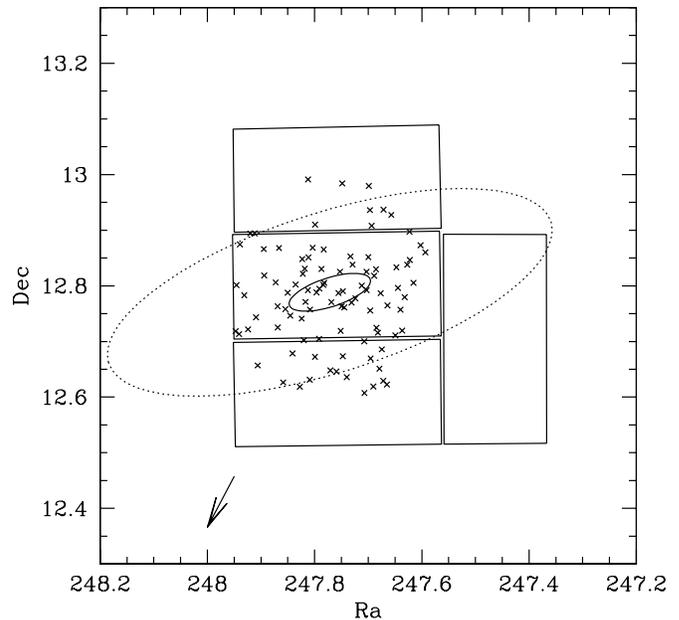}}
  \caption{Position on the sky of our observations. The coordinates
    are given in decimal degrees, epoch J2000. Central coordinates for 
    the galaxy are from \citet{2008ApJ...684.1075M}. The solid ellipse
    represents the core radius and the dotted ellipse the King profile limiting radius
    of the Hercules dSph galaxy as determined by
    \citet{2007ApJ...668L..43C}. Solid rectangles outline the four CCD
    chips in the WFC used for the photometric observations. $\times$
    marks the fibre positions for the FLAMES observations. Here we
    show the fibres positioned on the stars only. The arrow in the bottom left hand corner indicates the
    direction to the centre of the Milky Way.}
  \label{camera}
\end{figure}

Intermediate-band Str\"omgren $u$, $v$, $b$ and $y$ photometry was
obtained during six nights with the Wide Field Camera (WFC) on the
2.5-m Isaac Newton Telescope (INT) on La Palma. Of the six nights only
three provided useful data due to bad weather conditions. The WFC consists of 4 2k$\times$4k
CCDs. The CCDs have a pixel size of 13.5 microns corresponding to
0.33$\arcsec$ per pixel. Figure \ref{camera} shows the location and
dimensions of the galaxy on the sky and the positions of the CCD chips
of the WFC.

The observations are summarised in Table \ref{table:1} and all observations were centered at RA=$16^h 31^m 05^s$ and Dec=+$12^\circ 47' 18''$ \citep{2008ApJ...684.1075M}. Typical seeing during the three good nights was about 1.3 arcsec. \\

\begin{table}
\caption{Summary of the photometric observations obtained with the Isaac Newton Telescope.}              
\label{table:1}      
\centering                                      
\begin{tabular}{c c c c c}          
\hline\hline                        
Filter & 13 April 2007 & 14 April 2007  & 15 April 2007  & Total \\  
 & [$min$] & [$min$] & [$min$] & [$min$] \\  
\hline                                   
    $y$ & 1 x 30 & 1 x 30 & 1 x 30 & 90 \\     
    $b$ & 1 x 30 & 1 x 30 & 1 x 30 & 90 \\
    $v$ & 1 x 30 & 1 x 30 & 1 x 30 & 90 \\
    $u$ & 1 x 30 & 2 x 30 & 2 x 30 & 150 \\
\hline                                        
\end{tabular}
\begin{list}{}{}
\item[] Column 1 lists the filter. Column 2-4 list the number of 30 minutes exposures obtained for each filter during each of the three useful nights. Column 5 lists the total exposure time for each filter.
\end{list}
\end{table}

Multiple standard and extinction stars were observed each night. During the observations of the Hercules dSph galaxy, we observed 12 Str\"omgren standard stars chosen from the list in \citet{1988A&AS...73..225S} plus one star from \cite{1993A&AS..102...89O}, see Table \ref{table:std}. Two stars from the list of standard stars were observed several times during each night in order to sample the extinction for a large range of airmass. These stars will henceforth be referred to as extinction stars. \\
The observation of standard and extinction stars are used to find the zeropoint, extinction coefficients and colour terms, see Sect. \ref{cal}.

\begin{table}
\caption{Standard stars from \citet{1988A&AS...73..225S}, except HD107853 \citep{1993A&AS..102...89O}.}      
\label{table:std}    
\centering                          
\begin{tabular}{l c c c c c}          
\hline\hline                       
ID & Hip &   $V$   &   $(b-y)$   &   $m_{1}$   &   $c_{1}$   \\  
\hline                                  
HD 100363 & 56327 & 8.648 & 0.191 & 0.139 & 0.760 \\  
HD 107853 & ... & 9.100 & 0.321 & 0.157 & 0.472  \\ 
HD 108754 & 60956 & 9.006 & 0.435 & 0.217 & 0.254 \\ 
HD 120467 & 67487 & 8.147 & 0.728 & 0.757 & 0.088 \\  
HD 134439 & 74235 & 9.058 & 0.484 & 0.224 & 0.165 \\  
HD 138648 & 76203 & 8.137 & 0.504 & 0.358 & 0.290 \\ 
HD 149996 & 81461 & 8.495 & 0.396 & 0.164 & 0.305 \\  
HD 158226 & 85378 & 8.494 & 0.386 & 0.146 & 0.316 \\ 
DM -05 3063 & 51127 & 9.734 & 0.568 & 0.461 & 0.182 \\  
DM -08 4501 & 87062 & 10.591 & 0.452 & 0.032 & 0.274 \\  
DM -12 2669 & 43099 & 10.230 & 0.229 & 0.094 & 0.490 \\  
DM -13 3834 &  69232 & 10.685 & 0.415 & 0.098 & 0.183 \\  
DM -14 4454 & 81294   & 10.332 & 0.565 & 0.469 & 0.192 \\  
\hline                                           
\end{tabular}
\begin{list}{}{}
\item[] Column 1 and 2 list the star ID and Hipparcos number, respectively. Column 3 lists the standard values adopted for the magnitude (note $y\equiv V$) and column 4 to 6 the standard values adopted for the $(b-y)$, $m_{1}$ and $c_{1}$ indices, respectively. 
\end{list}
\end{table}

	\subsubsection{Reduction of the photometric observations}  \label{ccq}

The images for the Hercules dSph galaxy and the standard stars were reduced with the Wide Field Survey Pipeline provided by the Cambridge Astronomical Survey Unit \citep{2001NewAR..45..105I}. The processing operations applied to the images were de-biasing, trimming, flatfielding, astrometry and correction for non-linearity. 

	\subsubsection{Standard star photometry and establishing the photometric calibration}  \label{cal}
	
\begin{figure}
\resizebox{\hsize}{!}{\includegraphics{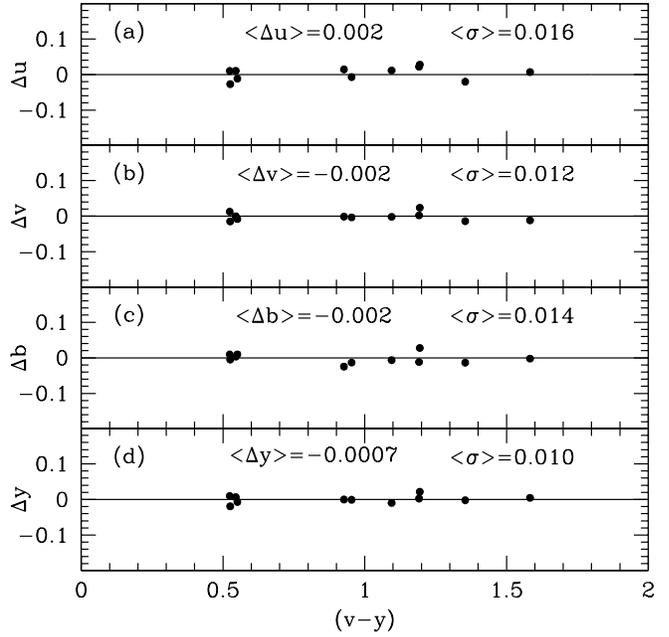}} 
  \caption{The residuals for the standard stars as a function of our
    final $(v-y)$ colour. The residuals are given in the sense "our
    observed value" - "the standard value". Mean differences and $\sigma$ as indicated.}
  \label{residual}
\end{figure}

We obtained aperture photometry for the standard and extinction stars using the task PHOT within the IRAF \footnote{IRAF is distributed by the National Optical Astronomy Observatories, which are operated by the Association of Universities for Research in Astronomy, Inc., under cooperative agreement with the National Science Foundation.} APPHOT package. The size of the aperture was determined individually for each star by plotting the measured flux as a function of increasing aperture size. The aperture at which the flux no longer increased was chosen as the aperture for that star (typically $4-5\times {\rm FWHM}$ of the stellar $psf$). This curve-of-growth is used in order to maximise the signal-to-noise ratio (S/N) while measuring as much flux as possible from the star \citep{1989PASP..101..616H}. \\ 

The measurements of the standard and extinction stars were used to establish the transformations needed to put our observations on the standard system of \cite{1993A&AS..102...89O}. Note that the \citet{1988A&AS...73..225S} stars are secondary standards in this system. See \cite{1995A&A...295..710O} for a discussion about the Olsen system as compared to the system established by \citet{1980ApJS...44..517B} and \citet{1994AJ....107.1577A}. \\
The first night out of the three useful nights did not give us reliable photometry for the standard stars (i.e. there was no well defined linear relation for the transformation to the standard system for that night), and this night was thus excluded from the calibration.

First, we derived preliminary extinction coefficients, $k_i$, and zeropoints, $z_i$, for each filter $i$ by solving the following equation with a least-square fit
\begin{equation} \label{calib1}
m_s=m_0+k_i \cdot X + z_i
\end{equation}
where $m$ is the magnitude of the star and $X$ is the airmass. The subscripts $s$, $0$ and $i$ designates the standard magnitude, the observed magnitude and the filter, respectively.
With preliminary zeropoints and extinction coefficients we then solved for the colour term, $a_i$, using a least-square fit
\begin{equation} \label{calib2}
m_s=m_0+a_i \cdot (v-y)_s + {z_i}'
\end{equation}
where ${z_i}^{'}$ is the residual from the linear fit. Note that here we use the data for both nights in order to make use of the colours for all standard stars.
With preliminary colour terms we can now solve the full equation to obtain better estimates of $z_i$ and $k_i$
\begin{equation} \label{calib3}
m_s=m_0+a_i \cdot (v-y)_s+{k_i}'' \cdot X + {z_i}''
\end{equation}
With these better estimates, Eq. (\ref{calib2}) is again solved and Eq. (\ref{calib2}) and (\ref{calib3}) are then iterated until convergence in $a$, $z$ and $k$ is achieved. Typically convergence is reached within 6 to 7 iterations. The final zeropoint is given by $z''$ in Eq. (\ref{calib3}) plus the residual zeropoint from Eq. (\ref{calib2}) from the last iteration. The extinction coefficients, colour terms, zeropoints and correlation coefficients between the uncertainty in $z$ and $k$, $\rho_{zk}$, for each filter and night are listed in Table \ref{table2}. We note that the uncertainty in zeropoint and extinction coefficient are strongly anti-correlated.

The $z$, $k$ and $a$ obtained were used to transform our observations onto the standard system of \cite{1993A&AS..102...89O}. The uncertainty for $z$, $k$ and $a$, and the correlation coefficient $\rho_{zk}$ were used to calculate the errors in the magnitudes, Sect. \ref{cer}.

\begin{table*}
\caption{Coefficients for Eq. (\ref{calib3}).}              
\label{table2}      
\centering                                     
\begin{tabular}{c c c c c c c c c c c c c}          
\hline\hline                        
Night & $k_y$ & $z_y$ & $a_y$ & $\rho_{zk}$ \\    
\hline                                   
14 April 2007 & --0.155 $\pm$ 0.021 & 23.009 $\pm$ 0.031 & 0.016 $\pm$ 0.005 & --0.92 \\
15 April 2007 & --0.142 $\pm$ 0.016 & 22.989 $\pm$ 0.023 & 0.016 $\pm$ 0.005 & --0.96 \\
\hline
Night & $k_b$ & $z_b$ & $a_b$ & $\rho_{zk}$ \\    
\hline
14 April 2007 & --0.240 $\pm$ 0.031 & 23.337 $\pm$ 0.043 & 0.009 $\pm$ 0.005 & --0.97 \\
15 April 2007 & --0.210 $\pm$ 0.036 & 23.297 $\pm$ 0.050 & 0.009 $\pm$ 0.005 & --0.96 \\
\hline
Night  & $k_v$ & $z_v$ & $a_v$ & $\rho_{zk}$ \\    
\hline
14 April 2007 & --0.349 $\pm$ 0.023 & 23.051 $\pm$ 0.034 & 0.050 $\pm$ 0.006 & --0.98 \\
15 April 2007 & --0.353 $\pm$ 0.023 & 23.048 $\pm$ 0.032 & 0.050 $\pm$ 0.006 & --0.96 \\
\hline
Night &  $k_u$ & $z_u$ & $a_u$ & $\rho_{zk}$ \\   
\hline
14 April 2007 & --0.582 $\pm$ 0.036 & 23.131 $\pm$  0.052 & 0.065 $\pm$ 0.007 & --0.98 \\
15 April 2007 & --0.569 $\pm$ 0.013 & 23.117 $\pm$  0.021 & 0.065 $\pm$ 0.007 & --0.96 \\
\hline                                             
\end{tabular}
\begin{list}{}{}
\item[] Columns 2 to 4 list the airmass extinction coefficients, $k_i$, zeropoints, $z_i$ and colour coefficients, $a_i$, for each filter as indicated with uncertainties $\sigma_k$, $\sigma_z$ and $\sigma_a$ respectively. Column 5 lists the correlation coefficient between the uncertainty in extinction coefficient, $\sigma_k$, and uncertainty in zeropoint, $\sigma_z$. This coefficient is denoted by $\rho_{zk}$.
\end{list}
\end{table*}

In Fig. \ref{residual} we show the residuals between our photometry and the standard values from \citet{1988A&AS...73..225S} and \cite{1993A&AS..102...89O}  as a function of our calibrated $(v-y)$ colour. We note that there are no trends with colour.

	\subsubsection{Photometry of the stars in the science images}  \label{c}

Instead of co-adding the science images, we did aperture photometry on each of the images separately. Co-adding the images would be difficult since the seeing varied from night to night and because of the necessity to apply the extinction correction for each night separately. As described later in this section the final flux for each star in each filter was calculated using a weighted-mean of the individual measurements. This enables us to do a more detailed study of the night to night quality. As a quality check we compared the flux for the brightest targets for each night with a mean flux, calculated for each object for all nights, to see if any of the images deviated in flux. None of the images deviated. From this we draw the conclusion that the calibration was consistent throughout the entire observing run.

Coordinate lists for the images were created using the task DAOFIND in
the APPHOT package in IRAF. To establish the coordinate lists we used
the best $y$ image, since the stars are brightest in this filter. We
then used this catalogue of coordinates for all the other images. We
used the aperture photometry task PHOT, within the APPHOT package, to
measure the flux for all objects on the images.

\paragraph{Aperture correction.}
When doing photometry on the science images we used a fixed aperture
of 5 pixels. Applying a  curve-of-growth we obtained, for each
individual image, the aperture correction out to $4 \times {\rm FWHM}$
of the $psf$. The aperture corrections were based on measurements of
many bright isolated stars, typically 20 stars per CCD. The aperture corrections were done in
flux-space and are listed in Table\,\ref{apcotab}.

\begin{table}
\caption{Aperture corrections.}         
\label{apcotab}  
\centering          
\begin{tabular}{l c c c c c}   
\hline\hline   
File name & Filter & CCD chip & Aperture correction  \\  
\hline
r556732  & $y$ & 1/2/3/4 & 1.265/1.266/1.278/1.266  \\
r556737  & $b$ & 1/2/3/4 & 1.273/1.262/1.274/1.256  \\
r556748  & $v$ & 1/2/3/4 & 1.195/1.191/1.164/1.168  \\
r556754  & $u$ & 1/2/3/4 & 1.171/1.161/1.135/1.148  \\
r556882  & $b$ & 1/2/3/4 & 1.269/1.264/1.259/1.240  \\
r556887  & $y$ & 1/2/3/4 & 1.162/1.166/1.162/1.155  \\
r556893  & $v$ & 1/2/3/4 & 1.268/1.264/1.223/1.238  \\
r556898  & $u$ & 1/2/3/4 & 1.378/1.366/1.312/1.343  \\
r556899  & $u$ & 1/2/3/4 & 1.384/1.362/1.316/1.352  \\
r556997  & $y$ & 1/2/3/4 & 1.254/1.261/1.254/1.244  \\
r557004  & $b$ & 1/2/3/4 & 1.212/1.203/1.204/1.188  \\
r557009  & $v$ & 1/2/3/4 & 1.382/1.382/1.336/1.356  \\
r557018  & $u$ & 1/2/3/4 & 1.209/1.198/1.160/1.185  \\
r557019  & $u$ & 1/2/3/4 & 1.335/1.322/1.282/1.308  \\
\hline
\end{tabular}
\begin{list}{}{}
\item[] Column 1 lists the file name for the image as named by the observing and archiving system on the Isaac Newton Telescope. Column 2 lists the filter. Column 3 lists the CCD chip number and column 4 the aperture correction for each CCD in the same order as in column 3.
\end{list}
\end{table}

\paragraph{Final magnitudes.}

Initial magnitudes were calculated for each object and night for every image and
calibrated for the airmass extinction and zeropoint using
Eq. (\ref{calib2}), but this time with subscript $s$ as our calibrated
magnitude and, as before, $0$ as the observed magnitude, with coefficients from
Table \ref{table2}.  Since the first night did not give us reliable
standard star photometry, and we thus have no calibration for that
night, we normalized the magnitudes from that night to the mean of the
magnitudes for the two following nights. \\

For all three nights, erroneous measurements returned from PHOT for the individual exposures were removed from the data set (i.e. the measurements for which {\tt sier, cier {\rm and} pier}$\neq$0, which are the error in sky fitting, centering algorithm and photometry, respectively).
The flux was then calculated for each star. The final flux, $\overline{F}$, was obtained by using a weighted-mean flux where the photometric errors returned from PHOT ({\tt merr}) were used as weights. The expression for the final flux is thus

\begin{equation} \label{wemean}
\overline{F}=\frac{\sum^n_{j=1}f_j/\sigma_j^2}{\sum^n_{j=1}1/\sigma_j^2}
\end{equation}
where the subscript $j$ is the exposure, $n$ is the total number of exposures, $f_j$ is the flux of the individual exposure and $\sigma_j$ the error ({\tt merr}).
These mean fluxes were then converted back to magnitudes and the colour terms were applied to get the final magnitudes. 
By definition, $y\equiv V$ \citep[e.g.][]{1983A&AS...54...55O}, and we will henceforth use $V$ instead of $y$ in figures and discussions.

	\subsubsection{Photometric errors}  \label{cer}
	
\begin{figure}
\resizebox{\hsize}{!}{\includegraphics{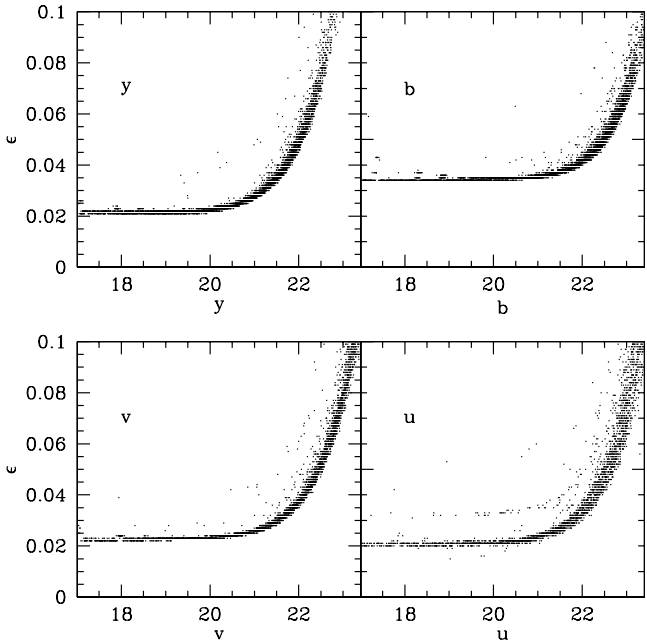}} 
\caption{Errors in the photometry for $y, b, v$ and $u$, as indicated. The calculation of the errors is described in Sect. \ref{cer}. The few stars with anomalously high errors at a given magnitude are also discussed there.}
\label{err}
\end{figure}

The errors in magnitude for the Str\"omgren photometry were calculated using a Monte Carlo Simulation, taking into account {\tt merr}, the uncertainty in zeropoint, extinction coefficient and colour term and the strong anti-correlation between the uncertainty in zeropoint and extinction coefficient. \\

We did this in the following way. For each object, a new magnitude was calculated
\begin{equation}
m_i=m_{0,i}+\Delta m
\end{equation}
where $m_0$ is the observed magnitude in filter $i$ and $\Delta m$ is a random number, drawn from a normal distribution with a mean of 0 and a variance of $\sigma^2_m$, where $\sigma_m$ is the error in the magnitude as returned from the PHOT task ({\tt merr}).
Additionally, for each object, a new zeropoint and extinction coefficient were calculated
\begin{equation}
z_i=z_{0,i}+\Delta z
\end{equation}
\begin{equation}
k_i=k_{0,i}+\Delta k
\end{equation}
where $z_{0,i}$ and $k_{0,i}$ are the zeropoint and extinction coefficient for filter $i$ used in the calibration( see Sect. \ref{cal}) and $\Delta z$ is a random number, drawn from a normal distribution with a mean of 0 and a variance of $\sigma^2_z$, see Table \ref{table2}. Since $\sigma_z$ and $\sigma_k$ are strongly anti-correlated, we take into account the correlation coefficient from the calibration when calculating the random number $\Delta k$ (see Sect. \ref{cal} and Table \ref{table2}). Finally, a new colour term was calculated
\begin{equation}
a_i=a_{0,i}+\Delta a
\end{equation}
where $a_{0,i}$ is the calculated colour term for filter $i$ and $\Delta a$ is a random number, drawn from a normal distribution with a mean of 0 and a variance of $\sigma^2_a$.

This process was then repeated 2000 times, generating a new set of $u$, $v$, $b$ and $y$ magnitudes in each iteration. As the final error for each magnitude we adopt the standard deviation of the magnitudes from the Monte Carlo simulation. This is calculated as follows
\begin{equation} \label{weerr}
\sigma_{\overline{m}}=\sqrt{\frac{1}{n-1}\sum^n_{j=1}{\left(m_j-\overline{m}\right)^2}}
\end{equation}
where the subscript $j$ is the Monte Carlo iteration number, and $\overline{m}$ the mean magnitude of the distribution.

Figure \ref{err} presents our final photometric errors,
$\epsilon$. 
The base-level error of  $\sim 0.02$ mag (0.035 for the $b$ filter) is due mainly to the errors in the photometric calibration, see Table \ref{table2} and the discussion above. The profile of the errors, i.e. increasing error with decreasing magnitude, is dominated by
{\tt merr} (i.e. photon statistics).
Some stars show a larger $\epsilon$ than the majority of
stars at that magnitude. This is most obvious in the $u$ filter and
is a statistical feature caused by the number of exposures that are
included for that star. The $u$ filter suffers most from
this since stars are generally a lot fainter in this filter, and
therefore more measurements are rejected due to errors in sky fitting,
centering and photometry. For example, the stars in $u$ with a larger
$\epsilon$ for a given magnitude have three or fewer individual measurements while
the stars in the main trend all have five measurements. The higher error in $b$ is caused 
by the higher uncertainty for the zeropoint in $b$ (compare Table \ref{table2}).

	\subsubsection{Stellar classification using SExtractor}  \label{se}
	
\begin{figure}
\resizebox{\hsize}{!}{\includegraphics{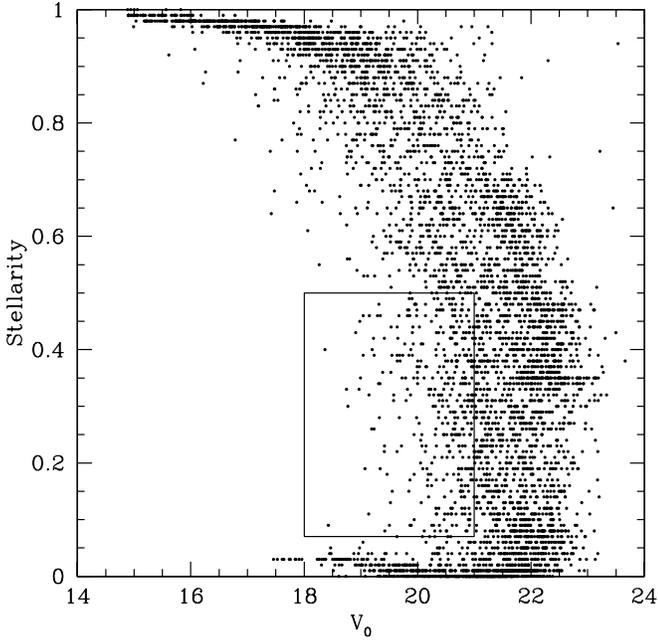}}
  \caption{SExtractor stellarity index ($sclass$) as a function of $V_0$. A star takes on values close to 1 and galaxies values close to 0. The rectangle indicates the area where we investigated each object individually (see Sect. \ref{se}).}
  \label{sclass}
\end{figure}

Contamination by background galaxies is a concern for our observations. As our images are uncrowded we can reach faint magnitudes with relative ease using aperture photometry. However, by using aperture photometry we have no information about the shape of the objects. In order to sort the stars from background galaxies we used SExtractor v2.5. SExtractor uses a tunable neural network trained on realistic simulated images to separate galaxies from stars in moderately crowded images \citep{1996A&AS..117..393B}. The SExtractor output, which is of interest to us, is the stellarity index, $sclass$; $sclass$ takes on values between 0 and 1, where 0 indicates a galaxy and 1 a star.
Following \citet{1996A&AS..117..393B}, objects with a stellarity index greater than or equal to 0.5 were identified as stars.

As can be seen in Fig. \ref{sclass}, SExtactor clearly removes galaxies at bright magnitudes but its ability to distinguish stars from galaxies diminishes at fainter magnitudes where the S/N is lower.
During our membership analysis (see Sect. \ref{stsel}) we investigated objects in a stellarity index range $0.07<sclass<0.5$ and magnitude range $18<V_0<21$ in order to make sure that we did not exclude any objects of interest based on the stellarity index alone. This area is marked with a solid line in Fig. \ref{sclass}.
We found two objects of interest, INT 34489 and INT43290 (see Sect. \ref{stsel} for a discussion of these two objects).

	\subsection{Spectroscopy}  \label{d}

Our spectroscopy was carried out using the multiobject spectrograph Fibre Large Array Multi Element Spectrograph (FLAMES) at the Very Large Telescope (VLT) on Paranal. Operated in Medusa fibre mode, this instrument allows for the observation of up to 130 targets at the same time \citep{2002Msngr.110....1P}. Figure \ref{camera} shows the fibre positions on the sky for the stellar targets. 21 additional fibres were dedicated to observing blank sky. We used the GIRAFFE/L8 grating, which provides a nominal spectral resolution of $R \sim 6500$ and a wavelength coverage from 821 nm to 940 nm, centred on the Ca {\sc ii} IR triplet lines in the spectral region around 860 nm.

\begin{table}
\caption{Summary of the spectroscopic observations with FLAMES.}
\label{table:3}  
\centering                     
\begin{tabular}{c c}     
\hline\hline          
Date & Exp. time [$min$] \\   
\hline                         
15 April 2007 & 45 \\  
15 April 2007 & 45 \\
10 May 2007 & 45 \\
17 May 2007 & 40 \\
21 June 2007 & 40 \\
21 June 2007 & 40 \\
21 June 2007 & 40 \\
\hline
Total Exp. Time & 295 \\
\hline
\end{tabular}
\begin{list}{}{}
\item[] Column 1 lists the date of observation and column 2 the exposure time.
\end{list}
\end{table}

	\subsubsection{Reduction of spectroscopic observations}

Initially, the FLAMES observations were reduced with the standard GIRAFFE pipeline version 2.2 \citep{2000SPIE.4008..467B}. This version of the pipeline, however, was not able to reduce a persistent glow on the CCD \citep{2008A&A...490..777L}. This glow then created an extra background that increases towards the red. The effect was very large and would have affected the equivalent width measurements.
Fortunately, during our work with these spectra a beta-version of the next version of the pipeline became available. The data were thus re-reduced with the GIRAFFE pipeline, version 2.5. This pipeline provides bias subtraction, flat fielding, dark-current subtraction, and accurate wavelength calibration from a ThAr lamp. It also solved the issue of the CCD glow.

	\subsubsection{Spectroscopic measurements}\label{specmea}

\begin{figure*}
  \resizebox{\hsize}{!}{\includegraphics{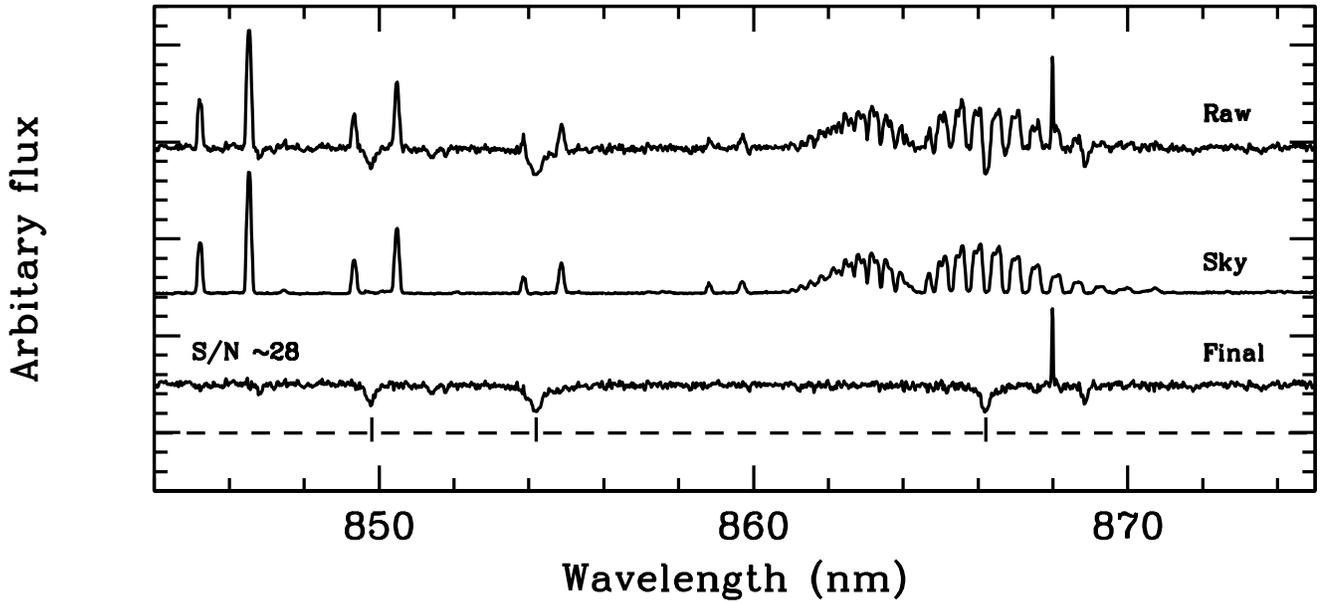}}
  \caption{Example showing the subtraction of sky-emission lines from the stellar spectra. At the top we show a raw, one-dimensional, spectrum for one of the brightest stars in our sample, $V=18.4$. Below that spectrum is the final spectrum of the sky constructed through the combination of all 21 sky fibres. At the bottom we show the stellar spectrum once the sky has been removed. The S/N for the final stellar spectrum is 28. The RMS in the sky spectrum is $\sim 1.1$ counts and $\sim 9.8$ counts in the stellar spectrum. The three Ca {\sc ii} IR triplet lines are indicated by vertical lines in the bottom spectrum. The dashed line indicates the position of zero intensity for the final spectrum.}
  \label{spec}
\end{figure*}

The 21 sky spectra were combined and subtracted from the object spectra with the task SKYSUB in the SPECRED package in IRAF. Figure \ref{spec} gives an example of the sky-subtraction process.

Finally, the object spectra from the individual frames were Doppler-shifted to the heliocentric rest frame and median-combined into the final one-dimensional spectrum. When combining the object spectra we used an average sigma clipping algorithm, rejecting measurements deviating by more than $3\, \sigma$, in order to remove cosmic rays.

Radial velocities were determined by a Fourier cross-correlation of the combined spectra against a synthetic template spectrum
using the IRAF task FXCOR. The template consisted of three Gaussian absorption lines at the positions of the Ca {\sc ii} IR triplet lines, with equivalent widths ($W$) representative for red giant stars. The radial velocities were determined from a Gaussian fit to the strongest correlation peak within a 300 km\,s$^{-1}$ window. The uncertainty in the measurement of the radial velocity was returned by FXCOR and is based on the Tonry-Davis R-value \citep{1979AJ.....84.1511T} (see Fig. \ref{sclass2}a and b). 

During the Fourier cross-correlation process we performed an ocular inspection of the quality of the spectra. Spectra of objects fainter than $V_0\sim21.3$ showed no clearly visible Ca {\sc ii} IR triplet lines and were thus removed from the sample. The S/N for these spectra was typically $\sim 4$ or less.

The equivalent widths, $W$, for the Ca {\sc ii} IR triplet lines were measured by fitting a Gaussian profile \citep{2004MNRAS.347..367C} to each of the three lines using the IRAF task SPLOT. From an ocular inspection of the spectra we found that the Gaussian profile fitted the Ca {\sc ii} triplet lines better than a Voigt profile.

	\subsection{Interstellar reddening} \label{red}

We corrected the photometric magnitudes for interstellar extinction using the dust maps by \citet{1998ApJ...500..525S}. This gives $E(B-V)=0.062$, in agreement with \citet{2008ApJ...688L..13K}.

\citet{2007ApJ...668L..43C} used a reddening of $E(B-V)=0.055$ with an uncertainty of 0.005 that represents the variation in reddening over the Large Binocular Telescope 23$\arcmin$ $\times$ 23$\arcmin$ field. 
In Sect. \ref{fehm1} we investigate how different values of $E(B-V)$ affect the estimated metallicities for the stars.

We used the \citet{1998ApJ...500..525S} relations to translate these extinction values into the Str\"omgren system. De-reddened magnitudes, colours and Str\"omgren indices will henceforth have the subscript $0$.

	\section{Results} \label{result}

	\subsection{Colour magnitude diagram in the direction of the Hercules dSph galaxy}  \label{e}

\begin{figure*}
  \resizebox{\hsize}{!}{\includegraphics{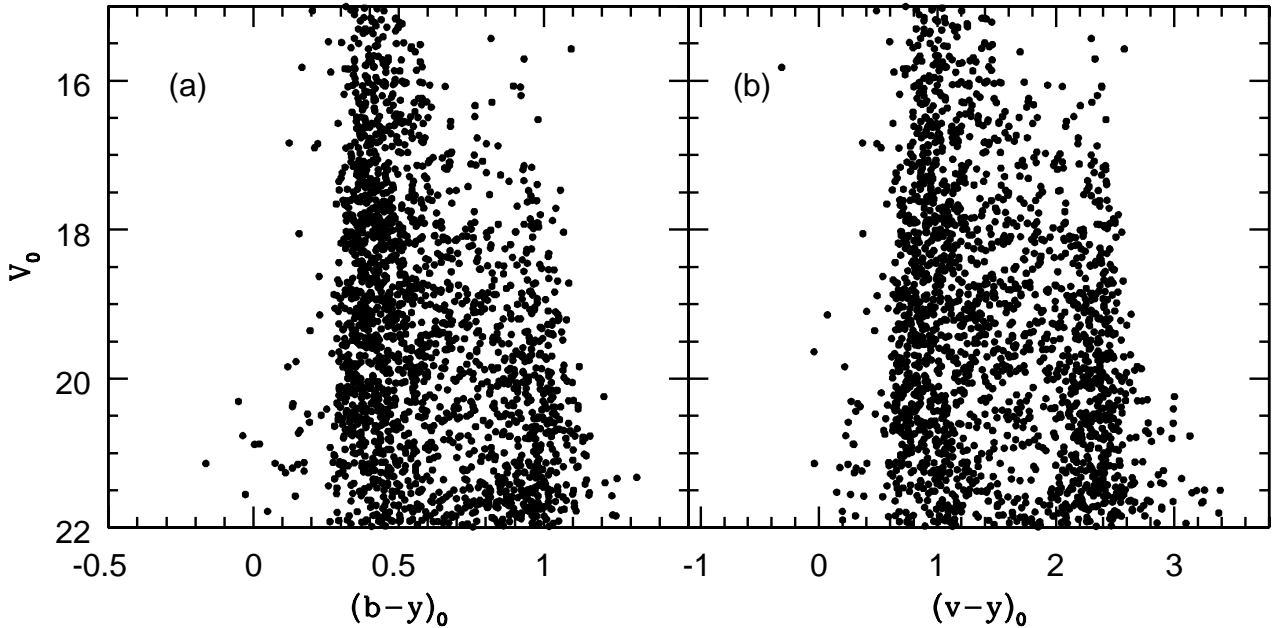}}
  \caption{Colour-magnitude diagram centred on the Hercules dSph galaxy. {\bf (a)} $V_0$ vs. $(b-y)_0$. {\bf (b)} $V_0$ vs. $(v-y)_0$. Only objects with a stellarity index $sclass\geq 0.5$ are shown (see Sect. \ref{se}).}
   \label{cmd4}
\end{figure*}

Figure \ref{cmd4} presents our colour magnitude diagrams in the direction towards the Hercules dSph galaxy. The horizontal branch (HB) is seen at $V_0 \approx 21.2$ in Fig. \ref{cmd4}a. A large population of foreground stars can also be seen with a cut-off at $(b-y)_0\approx 0.3$, associated with the blue limit of the turnoff stars in the Milky Way disk and halo. 
The RGB of the Hercules dSph galaxy cannot easily be seen due to the heavy contamination by foreground dwarf stars.

	\subsection{Radial velocities}  \label{ee}
	
\begin{figure*}
\resizebox{\hsize}{!}{\includegraphics{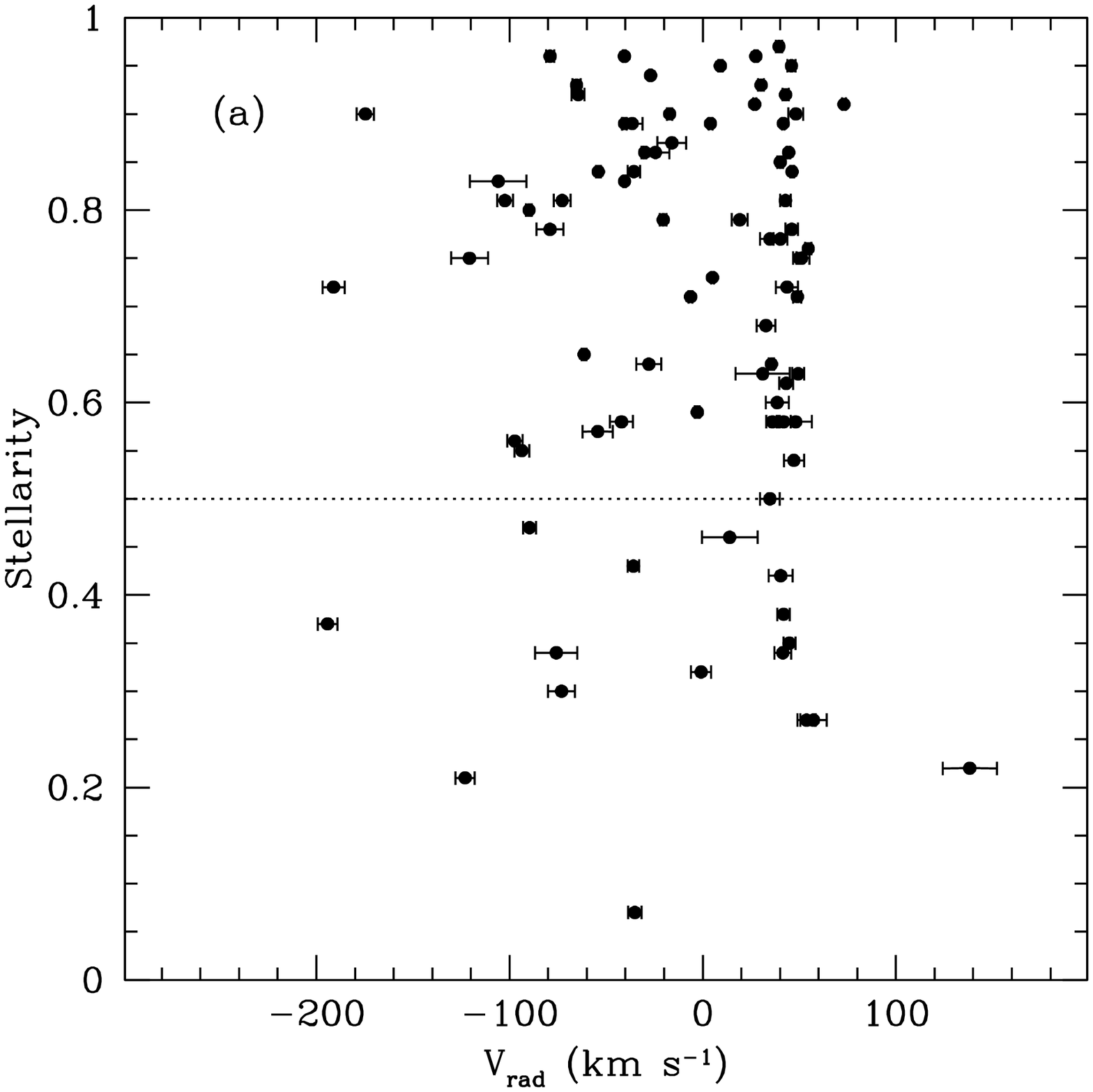}\includegraphics{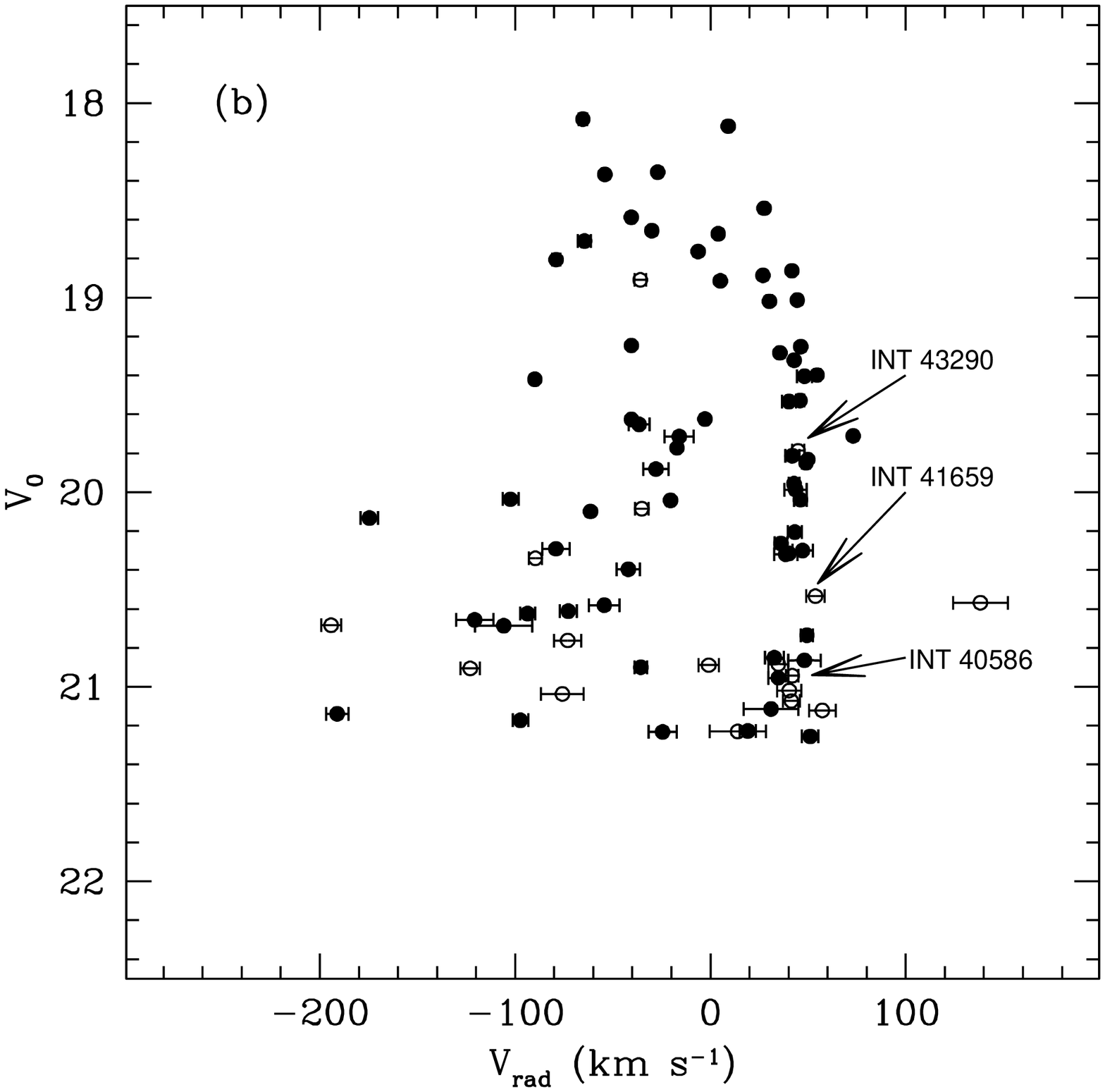}} 
  \caption{{\bf (a)} SExtractor stellarity index vs. radial velocity for the objects observed with FLAMES. The error-bars represent the error in $V_{rad}$ as returned by the task FXCOR. The dotted line indicates stellarity index 0.5. {\bf (b)} $V_0$  vs. radial velocity. $\bullet$ indicates objects with a stellarity index greater than 0.5. The error-bars represent the error in $V_{rad}$ as returned by the task FXCOR. $\circ$ indicates objects with a stellarity index lower than 0.5. Three stars are identified with their INT numbers. These stars are discussed in Sect. \ref{ee} and \ref{stsel}.}
  \label{sclass2}
\end{figure*}

\begin{figure}
\resizebox{\hsize}{!}{\includegraphics{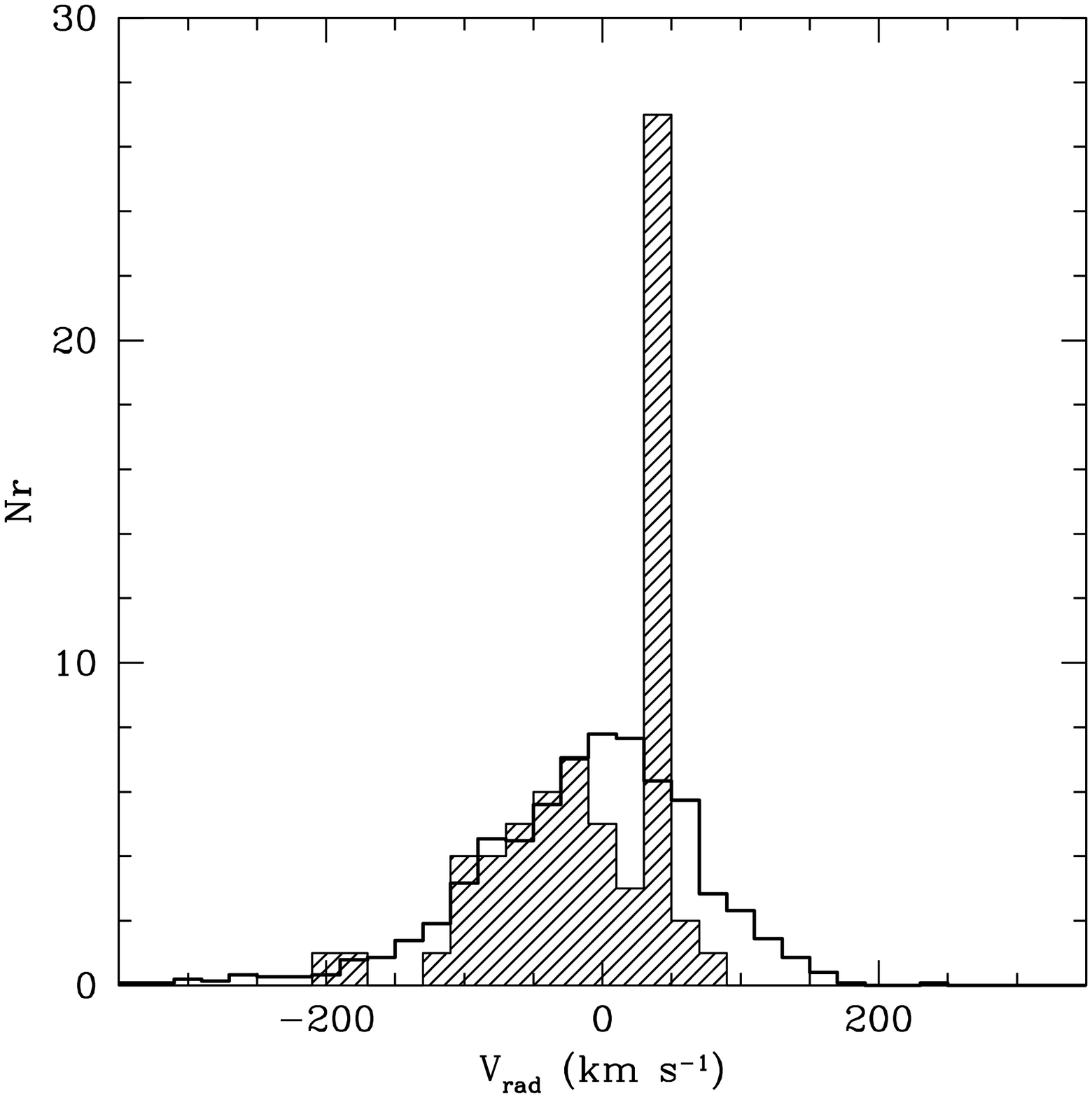}}
  \caption{Distributions of radial velocities. The solid histogram shows the distribution of velocities from a Besan\c{c}on model in the direction of the Hercules dSph galaxy. The shaded histogram shows the distribution of radial velocities for our FLAMES observations. Only objects with $sclass>0.5$ are included (see Sect. \ref{se}). Note that the area of the histogram based on the Besan\c{c}on model has been normalized to cover the same area as the FLAMES histogram.}
  \label{besancon}
\end{figure}

Figure \ref{sclass2}a shows the stellarity index as a function of the derived radial velocity for the objects observed with FLAMES. Objects with a stellarity index less than 0.5 are excluded from the sample (see Sect. \ref{se}). We investigated the objects with a stellarity index less than 0.5 and a velocity close to the systemic velocity and found that all of these objects are fainter than $V_0=21$ except three objects, INT 43290, INT 40586 and INT 41659. INT 43290 lies on top of a background galaxy and is therefore contaminated. INT 40586 and INT 41659 have a questionable surface distribution on the CCD. We flag them as blends and possible double-star systems and identify them in Fig. \ref{sclass2}b.

Fig. \ref{sclass2}b shows the magnitude of the stars vs. the derived radial velocities. The error-bars on the radial velocities are from the Tonry-Davis R-value \citep{1979AJ.....84.1511T} estimates. As expected, errors in the velocities correlate with magnitude such that fainter stars have larger errors.

In Fig. \ref{besancon} we compare the distribution of our radial
velocities in the direction of the Hercules dSph galaxy with the
predictions of the Besan\c{c}on model \citep{2003A&A...409..523R} with the same area on the sky as that covered by our observations. The
colour and magnitude range for the distribution of stars from the
Besan\c{c}on model is the same as spanned by the FLAMES targets. As
can be seen, also in this respect the Hercules dSph galaxy suffers from
heavy foreground contamination.
We note that the observed velocity distribution is not centred on the field star distribution predicted by the model, but is shifted by about $-25 {\rm \, km\, s^{-1}}$. Moreover, disregarding the peak created by the Hercules members, the observed distribution is narrower than the model prediction. This may point to some problem with the model; however,
a full investigation of the origin of this difference is beyond the scope of the present study.

	\section{Finding the giant and horizontal branch stars in the Hercules dSph galaxy} \label{findrgb}

	\subsection{The ability of the Str\"omgren $c_1$ index to identify RGB stars}  \label{sec1}
	
\begin{figure}
\resizebox{\hsize}{!}{\includegraphics{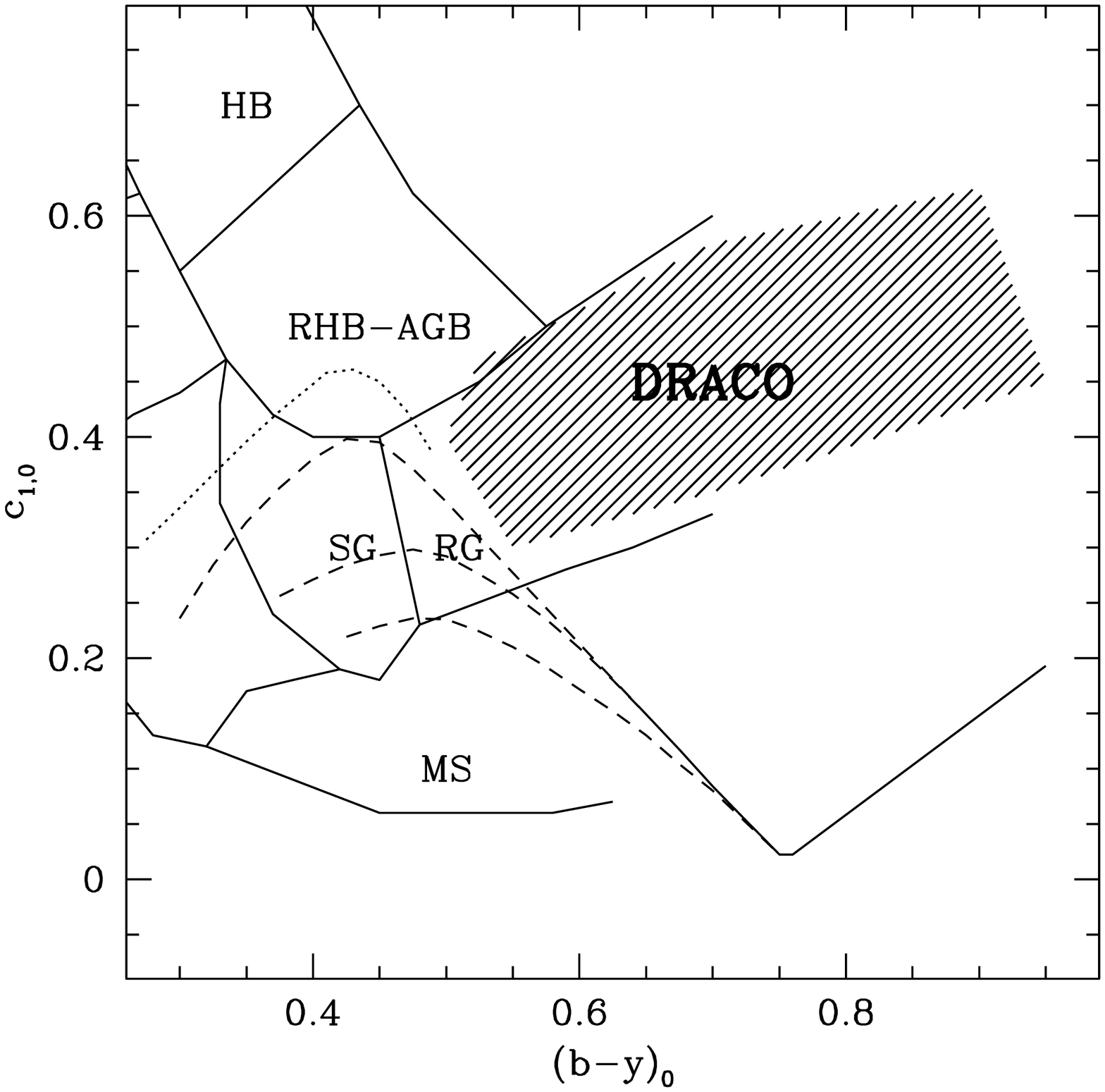}}
  \caption{$c_{1,0}$ vs. $(b-y)_0$ diagram with regions occupied by stars at various evolutionary stages as defined in \citet{2004A&A...422..527S}. MS: the main sequence; SG: sub-giant stars; RG: red giant stars; RHB-AGB: the red horizontal-branch-asymptotic-giant-branch transition; HB: the horizontal branch. The shaded region marked DRACO indicates the region occupied by the RGB stars in the Draco dSph galaxy \citep{2007A&A...465..357F}. The dashed lines indicate dwarf star sequences for different metallicities, ${\rm[Fe/H]}=0.45$, $-0.05$ and $-1$ from top to bottom and the dotted line marks the upper envelope for dwarf stars (all lines from \'{A}rnad\'{o}ttir et al. in preparation).}
  \label{schx}
\end{figure}

\begin{figure}
  \resizebox{\hsize}{!}{\includegraphics{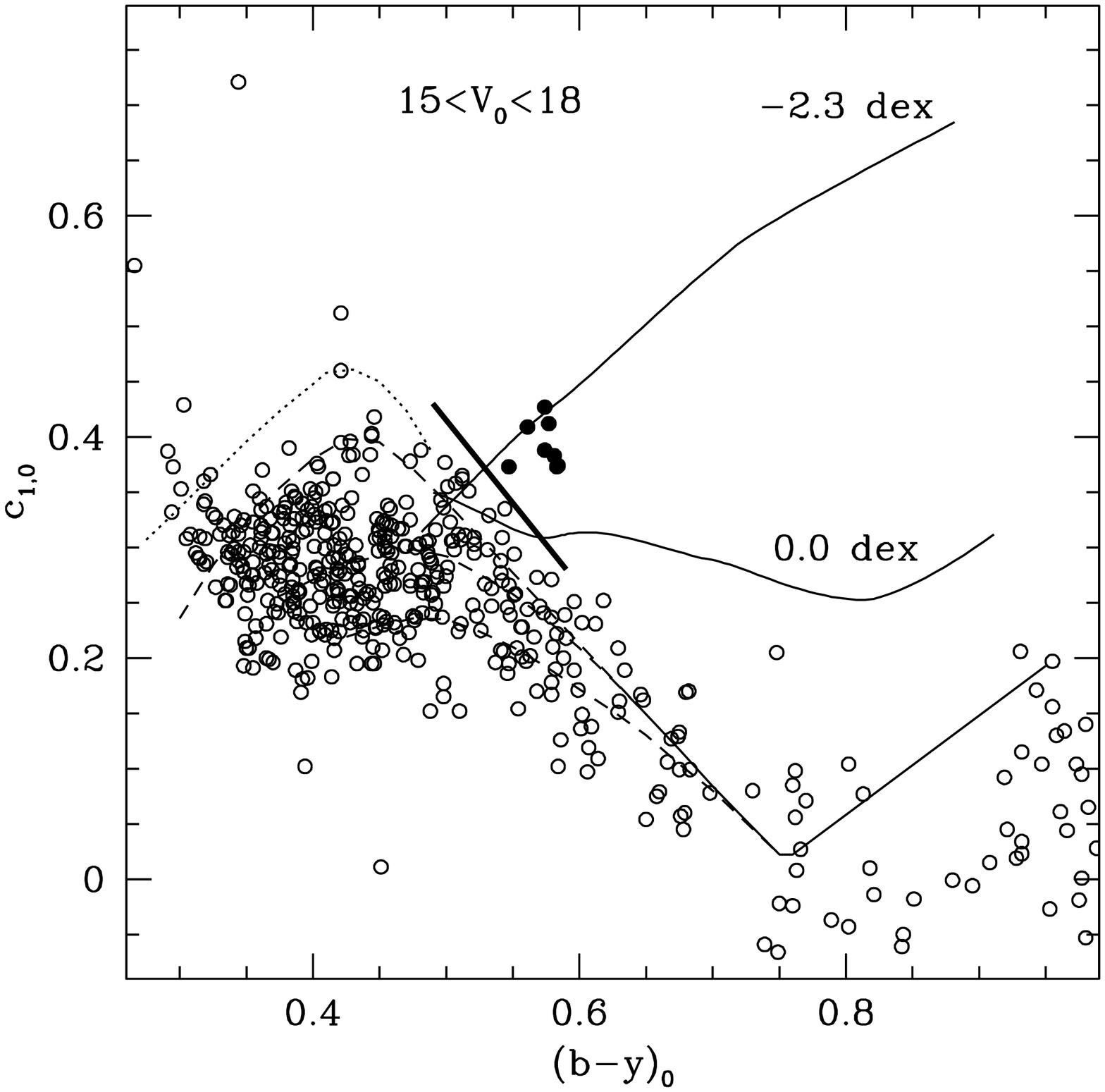}}
  \caption{$c_{1,0}$ vs. $(b-y)_0$ for stars, $\circ$, in the
    direction of the Hercules dSph galaxy. Only stars in the magnitude
    range $15<V_0<18$ are shown. $\bullet$ mark stars that fall
    within the RGB region as defined by the isochrones
    by \citet{2006ApJS..162..375V} with colour transformations
    by \citet{2004AJ....127.1227C}.
    Metallicities as indicated. The thin dashed lines
    indicates dwarf star sequences for different metallicities,
    ${\rm[Fe/H]}=0.45$, $-0.05$ and $-1$ top to bottom and the dotted line
    marks the upper envelope for dwarf stars (all lines from \'{A}rnad\'{o}ttir et al. in preparation). The thick line marks the
    empirically determined limit for the foreground contamination (see Sect. \ref{sec1}).}
  \label{plotc1bright}
\end{figure}

As Figs. \ref{cmd4} and \ref{besancon} show, the line of sight towards the Hercules dSph galaxy is heavily contaminated with foreground stars, making it impossible to determine membership from the colour-magnitude diagram alone. Even when radial
velocities are added the selection remains uncertain, because the mean velocity of the Hercules dSph galaxy coincides 
with the velocity of the (thick) disk. \\
The $c_1$ index in the Str\"omgren system gives us the ability to disentangle the RGB and HB stars in a dSph galaxy from the foreground dwarf stars.
The $c_1$ index is a measure of the Balmer discontinuity in a stellar spectrum and is defined as 
\begin{equation}
c_1=(u-v)-(v-b)
\label{c_1}
\end{equation}
The strength of the Balmer discontinuity depends on the evolutionary
stage of the star. Stars in a plot of $c_{1,0}$ vs. $(b-y)_0$ will
therefore occupy different regions depending on their evolutionary
stage. Figure \ref{schx} shows which regions are occupied by stars at
different evolutionary stages. This classification is adopted from \citet{2004A&A...422..527S}. We also show the region occupied by the
RGB stars in the Draco dSph galaxy \citep{2007A&A...465..357F}.

\citet{2004A&A...422..527S} were mainly concerned with high velocity
dwarf stars and to a lesser extent interested in the redder dwarf and
RGB stars. As the high-velocity halo stars that they studied tend to
be fairly blue we will use tracings for dwarf stars from
\'{A}rnad\'{o}ttir et al. (in preparation), and isochrones for RGB stars by
\citet{2006ApJS..162..375V} with colour transformations by \citet{2004AJ....127.1227C}, in the $c_{1,0}$
vs. $(b-y)_0$ diagram to define the dwarf and giant star
regions also in the red. The dwarf sequences in \'{A}rnad\'{o}ttir et
al. (in preparation) provide an extension of the preliminary dwarf relation from \citet{1984A&AS...57..443O}. The
major difference between the preliminary relation by \citet{1984A&AS...57..443O} and the
new relations is that the new relations are functions of
metallicity. In Fig. \ref{schx} we show dwarf sequences for three
different metallicities. These sequences, in accordance with
 \citet{1984A&AS...57..443O}, trace the
lower envelope for the dwarf stars (for $(b-y)$ less than about
0.55). Note that for redder colours ($(b-y)$ larger than about 0.55)
all dwarf sequences merge and form a single line that traces the mean
values of the colours.
By studying dwarf stars from \citet{1993A&AS..102...89O,1994A&AS..104..429O,1994A&AS..106..257O} in a $c_{1,0}$ vs. $(b-y)_0$ diagram it is possible to define an upper envelope for the region 
occupied by foreground dwarf stars.
\'{A}rnad\'{o}ttir et al. (in preparation) define such an upper envelope.
We include this in our plots henceforth.

To define the RGB region we use two isochrones with [Fe/H]$=-2.3$ and [Fe/H]$=0.0$ by \citet{2006ApJS..162..375V} and colour transformations by \citet{2004AJ....127.1227C}, see Fig. \ref{plotc1bright}.

As can be seen from Fig. \ref{plotc1bright}, for giant stars, the $c_{\rm 1}$ index has a clear metallicity
dependence. This is more prononounced for the reddest colours (i.e. the tip of the RGB).
However, in spite of this, this index still provides a strong discriminant between giant and
dwarf stars for cooler stars. Note that the dwarf sequence is not metallicity dependent
at these colours.

\paragraph{The blue limit for membership determination.}
Since in the $c_{1,0}$ vs. $(b-y)_0$ plane the dwarf and RGB stellar sequences converge around $(b-y)_0 \sim 0.5$ we need to identify a blue limit for stars that we identify as RGB stars. In Fig. \ref{plotc1bright} we show the $c_{1,0}$ vs. $(b-y)_0$ for stars in the magnitude range $15<V_0<18$. Stars fainter than $V_0=15$ are not saturated on the images and, based on our membership-determination in the next section, this magnitude range is bright enough not to contain any RGB stars in the Hercules dSph galaxy. We define a line that follows the upper envelope of observed dwarf stars to separate the RGB stars from the dwarf stars in order to safely exclude any dwarf stars. This line is somewhat higher in $c_{1,0}$ at a given $(b-y)_0$ than the tracing from \'{A}rnad\'{o}ttir et al. (in preparation). Our selection of RGB stars thus has a blue limit that is somewhat colour-dependent. No object bluer than this limit will be considered as a RGB star since they have a high probability of belonging to the foreground dwarf contamination. \\ 

In Fig. \ref{plotc1bright} we see 8 stars within the RGB region. A Besan\c{c}on model in the direction of the Hercules dSph galaxy gives $15$ RGB stars in the given magnitude and colour range. The conclusion is therefore that these stars are most likely foreground RGB stars belonging to the thick disc of the Milky Way.

	\subsection{Membership based on Str\"omgren photometry.} \label{stsel}

\begin{figure}
\resizebox{\hsize}{!}{\includegraphics{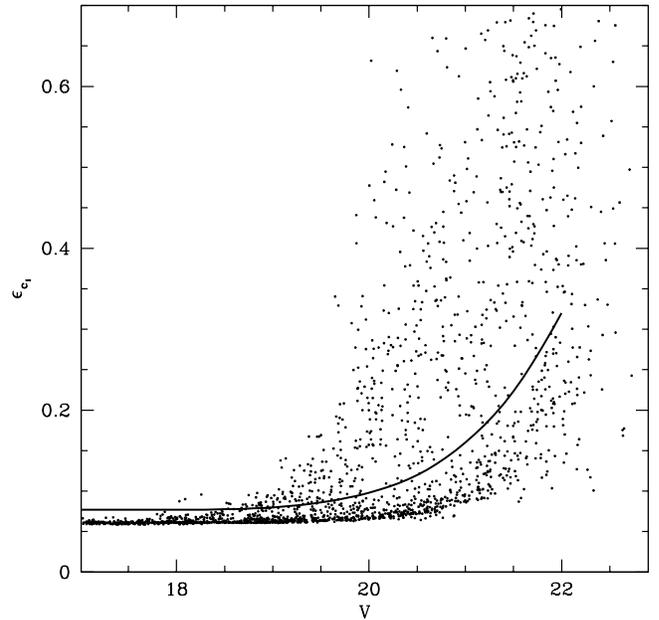}} 
\caption{Errors in the Str\"omgren colour index $c_1$ (see Sect. \ref{sec1}). The solid line indicates a spline function fitted to fall just above the trend with smaller errors. This line is used in Sect. \ref{stsel} and Fig. \ref{plot3} and \ref{plotc1err} where we define the RGB stars.}
\label{err2}
\end{figure}

\begin{figure*}
  \resizebox{18cm}{!}{\includegraphics{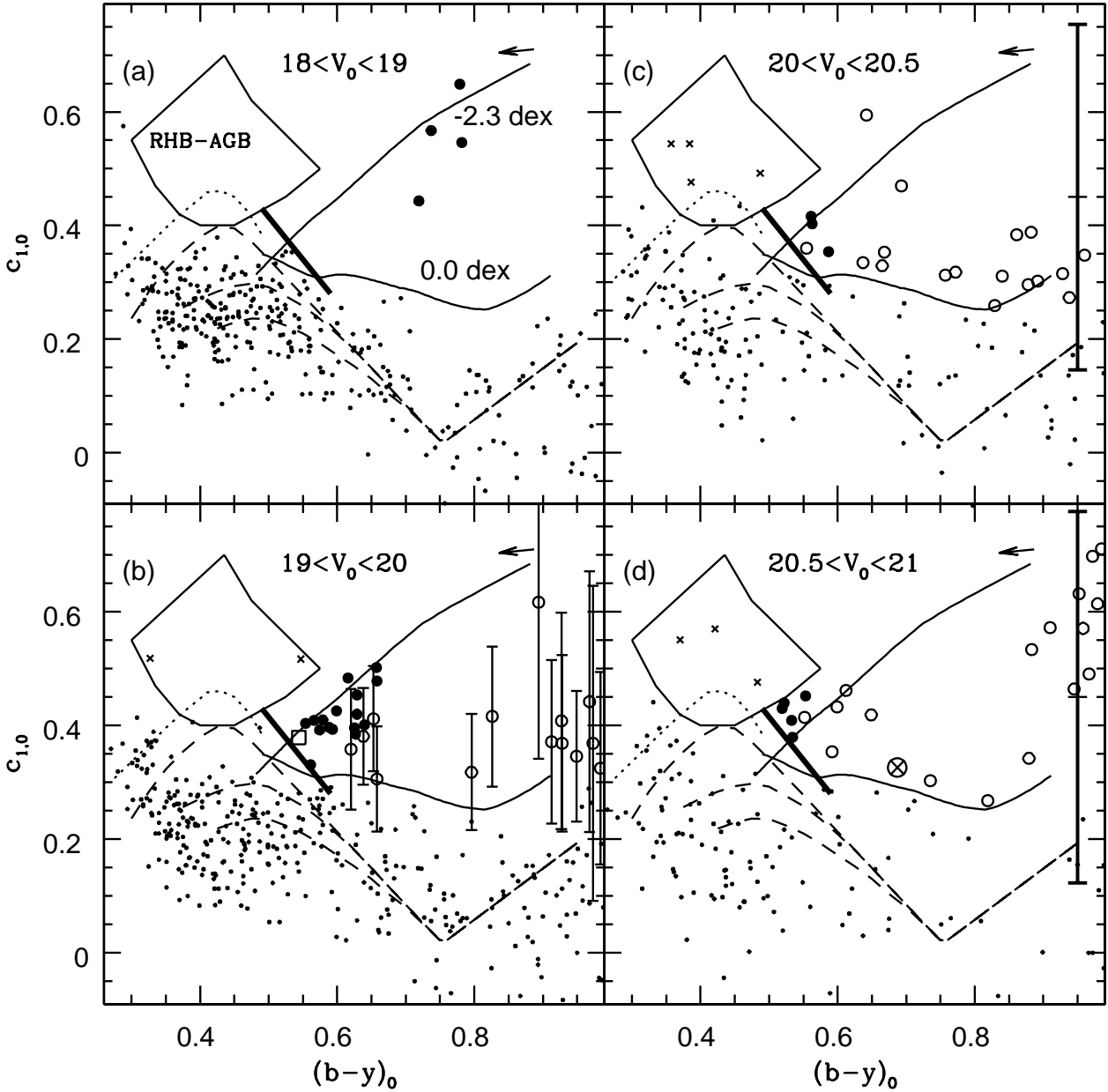}}
  \caption{$c_{1,0}$ vs. $(b-y)_0$ for stars in the direction
    of the Hercules dSph galaxy. The range of
    magnitude is indicated at the top of each panel. Dots are
    stars within the given magnitude range. $\bullet$ are stars that
    fall within the RGB region and have errors less than the line
    defined in Fig. \ref{err2}. These are members based on their evolutionary stage.
    $\circ$ are stars that fall within the RGB and have errors larger than the line
    defined in Fig. \ref{err2}. The star marked with $\otimes$ in
    (d) is flagged as a non-member due to its position in the
    $\epsilon_{c_1}$ vs. $(b-y)_0$ plane, see Fig. \ref{plotc1err}.
    The star marked with a $\Box$ in (c) is a 
    foreground RGB star (see Sect. \ref{stsel}). The error-bars in
    (b) represent the error in $c_1$ and are only displayed for stars
    that fall above the line defined in Fig. \ref{err2}. The thick
    error-bars to the right in (c) and (d) represent the mean error in
    $c_1$ for stars that fall above the line defined in
    Fig. \ref{err2} ($\circ$). $\times$ marks stars that fall in
    the RHB-AGB region (see Sect. \ref{hb}). The arrow in the
    top right corner in each panel indicates the magnitude and direction of the
    de-reddening applied to the data (see Sect. \ref{red}). 
    The solid lines indicate isochrones for RGB stars
    by \citet{2006ApJS..162..375V} with colour transformations
    by \citet{2004AJ....127.1227C}. Their metallicities are indicated in (a).
    The dashed lines indicate dwarf star sequences for different metallicities, ${\rm[Fe/H]}=0.45$, $-0.05$ and $-1$ from top to bottom and the dotted line marks the upper envelope for dwarf stars (all lines from \'{A}rnad\'{o}ttir et al. in preparation). The thick line marks the
    empirically determined limit for the foreground contamination (see Sect. \ref{sec1}).}
   \label{plot3}
\end{figure*}

\begin{figure}
  \resizebox{\hsize}{!}{\includegraphics{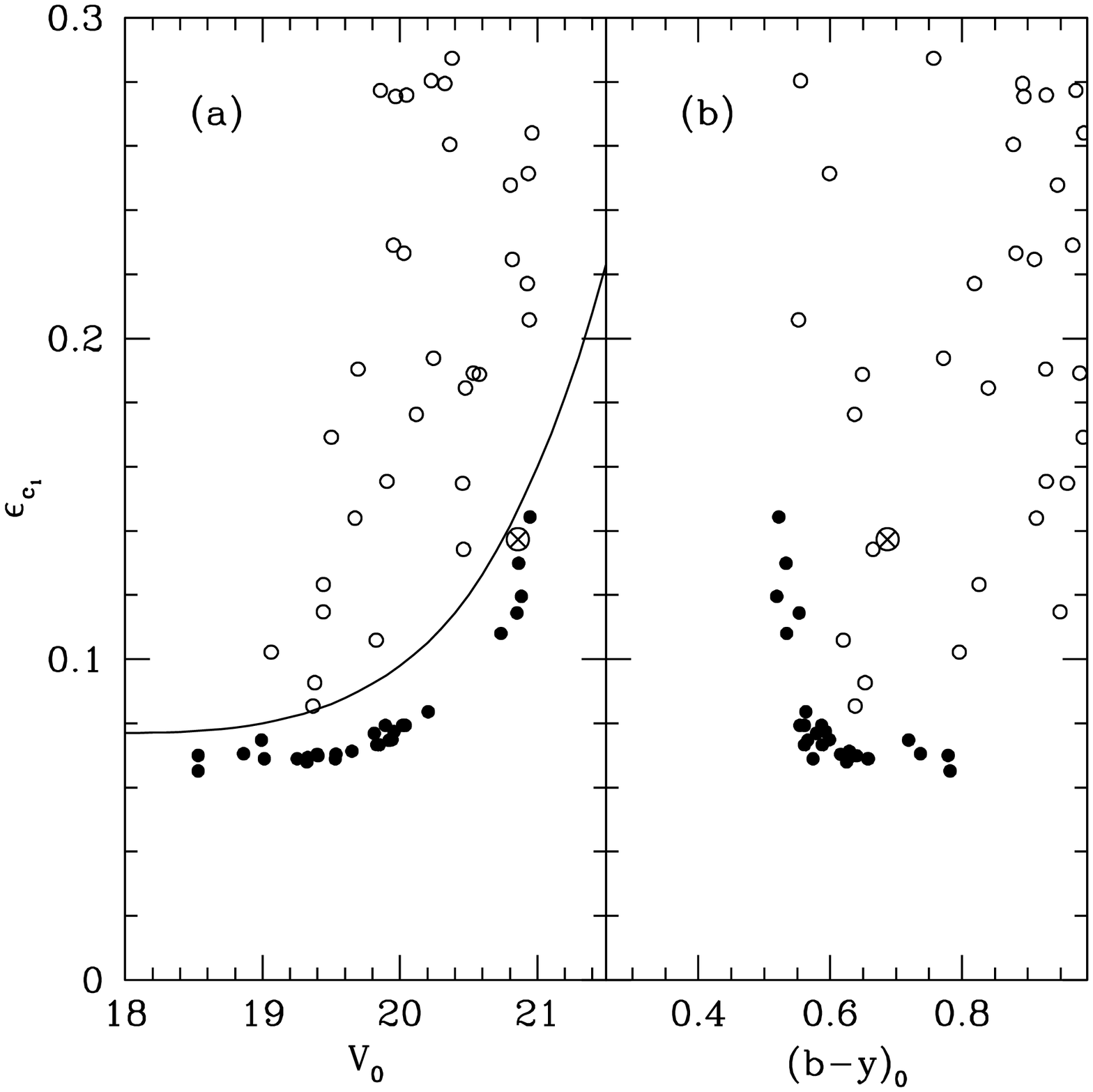}}
  \caption{$\epsilon_{c_1}$ vs. $V_0$ and $(b-y)_0$ for the stars in the RGB region identified in Fig. \ref{plot3}. Symbols are the same as in Fig. \ref{plot3} and the line in {\bf (a)} is the same as in Fig. \ref{err2}.}
  \label{plotc1err}
\end{figure}

We study the $c_{1,0}$ vs. $(b-y)_0$ in magnitude bins rather than the entire sample of all stars all at once. With this approach the RGB region is easier to track.
Figure \ref{err2} shows the errors in $c_{1,0}$ as a function of magnitude. We note the large errors in $c_1$ for
fainter stars. These errors are due to the higher errors in $u$ at any given magnitude. Note that in terms of errors
(see Fig. \ref{err}), $V=20$ corresponds to $u \sim 22.5$ for our RGB stars.
In order to exclude stars with large errors, we reject all stars with errors higher than the median error, plus 0.015 mag,
at a given magnitude (as indicated by the solid line in Fig. \ref{err2}).
We will now use this function in our classification of members and potential members of the Hercules dSph galaxy.

Figure \ref{plot3}a shows a clear and simple separation between the foreground dwarf stars and the four stars that fall in the RGB region. These four RGB stars all have errors in $c_{1,0}$ less than the line fitted in Fig. \ref{err2}.
Figure \ref{plot3}b-d show the three remaining magnitude bins. 
We also note that 9 stars fall within the RHB-AGB region, they
are discussed in Sect. \ref{hb}.

In Fig. \ref{plotc1err} we show $\epsilon_{c_1}$ vs. $V_0$ and $(b-y)_0$ for all the stars falling in the RGB region.

In total we find 29 member stars in the RGB region, and 9 in the RHB-AGB region. They are listed in Table \ref{photlist}.

\paragraph{Objects with a stellarity index lower than 0.5}
As mentioned in Sect. \ref{se} we found two objects of interest, based on their position in the $c_{1,0}$ vs. $(b-y)_0$ diagram, with a stellarity index lower than 0.5.
INT 34489 at $V_0=20.27$ has $sclass=0.3$ and is located right at the edge of the blue limit.
An ocular inspection of the object indicates that it may be a star, but it is slightly elongated on the CCD.
INT 43290 at $V_0=19.79$ has $sclass=0.35$. In Sect. \ref{ee} we found that this object lies on top of a galaxy and is therefore contaminated.
Given the higher uncertainty of these two objects we do not include them in further analysis. The stars are listed in Table \ref{lowsclass}.

\begin{table}
\caption{Objects with a stellarity index lower than 0.5, a radial velocity much lower than the mean velocity of the galaxy or $V_0>21$. }   
\begin{minipage}[t]{\columnwidth}
\label{lowsclass}
\centering                             
\begin{tabular}{l c c c c}          
\hline\hline                       
ID  & RA(2000) & DEC(2000) & $V_0$ & $V_{rad}$ \\    
\hline                                   
34489                                                       & 247.66907 & 13.07061 &  $20.27  \pm  0.02$ &   ...    \\
43290\footnote{On top of a galaxy}    & 247.63172 & 12.77969 &  $19.79  \pm  0.02$ &   $44.9  \pm    3.1$ \\
42668                                                       & 247.69964 & 12.85175 & $ 19.62  \pm  0.02 $ &   $-2.8  \pm     1.3$ \\
\hline
\end{tabular}
\end{minipage}
\begin{list}{}{}
\item[] Column 1 lists the INT ID. Column 2 and 3 list the coordinates. Column 4 lists the $V_0$ magnitude and its associated error. Column 5 lists the radial velocity and its associated error, both in km ${\rm s}^{-1}$.
\end{list}
\end{table}

\paragraph{A possible foreground RGB star}
In Fig. \ref{plot3}b we found one star, INT 42668, that fall within the
RGB region but it has a radial velocity of $V_{rad}=-2.8$\,km\,s$^{-1}$
(see Sect. \ref{vrmemb}) and ${\rm [Fe/H]_{Cal}}=-0.5$ (see
Sect. \ref{fehm1}). Since it has a radial velocity much lower than the mean velocity of the Hercules dSph galaxy, we identify this star as a likely foreground RGB star, thus removing it from the sample of member RGB stars. The star is listed in Table \ref{lowsclass}.

	\subsection{Finding the HB and AGB of the Hercules dSph galaxy}  \label{hb}
	
\begin{figure*}
  \resizebox{\hsize}{!}{\includegraphics{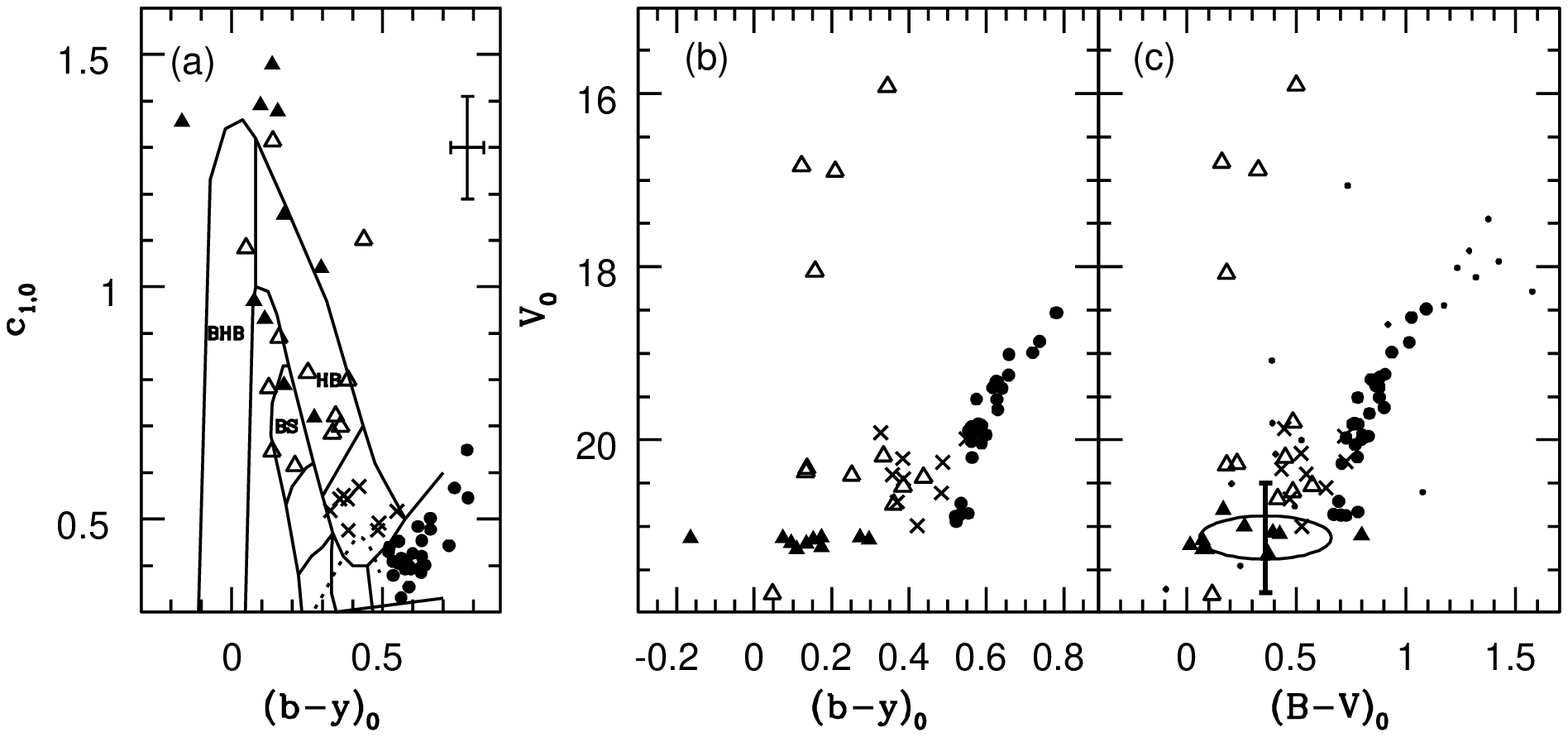}}
\caption{{\bf (a)} $c_{1,0}$ vs. $(b-y)_0$ for the stars in the
  direction of the Hercules dSph galaxy. BHB: the blue horizontal
  branch; HB: the remainder of the horizontal branch; BS: the blue
  stragglers \citep{2004A&A...422..527S}. The dotted line marks the
  upper envelope for dwarf stars. $\bullet$ are stars identified as
  RGB members and $\times$ are stars identified as RHB-AGB stars based
  on their evolutionary stage (compare Fig. \ref{schx} and \ref{plot3}). $\bigtriangleup$ are stars that fall on or are near the
  HB region but with a large offset to the expected HB
  magnitude. Filled triangles are stars that fall on or near the
  HB region and with the expected HB magnitude, they are thus
  identified as HB member stars. Error-bars in the top right corner
  are the mean $\epsilon_{c_{1,0}}$ and $\epsilon_{(b-y)_0}$ for the
  stars identified as HB stars. {\bf (b)} Colour-magnitude
  diagram in the Str\"omgren system for the stars identified in {\bf (a)}. {\bf (c)} Colour-magnitude diagram in the Johson-Cousin UBVRI system for the stars identified in {\bf (a)}. The solid ellipse
  outlines the distribution of variable stars in the Draco dSph galaxy
  from \citet{2004AJ....127..861B}, normalized in $V_0$ to fit our HB
  and $\cdot$ marks variable stars from \citet{2004AJ....127..861B}
  that fall outside the ellipse. The error bar on the ellipse
  corresponds to the mean amplitude of the variable stars used to define the
  ellipse.}
\label{plot28}
\end{figure*}

In Fig. \ref{plot3} we identified RGB members of the Hercules dSph
galaxy. Additionally, we found 9 RHB-AGB stars. However, it is possible
to use the Str\"omgren photometry to further explore the RHB-AGB and
blue HB stars (BHB). Figure \ref{plot28}a shows the BHB, HB and blue
straggler (BS) regions in the $c_1$ vs. $(b-y)$ plane using the areas defined in \citet{2004A&A...422..527S}.

In Fig. \ref{plot28}a and b we show the previously identified RGB and
RHB-AGB stars. In addition we show all stars bluewards of
$(b-y)_0=0.5$ with an $\epsilon_{c_1}$ less than the function in
Fig. \ref{err2}. We further divide these stars into two sets according
to their $V$ magnitude. The HB of the Hercules dSph galaxy is roughly
at $V_0=21.2$ (compare Fig. \ref{cmd4} and
\citet{2007ApJ...668L..43C}). We therefore define a box with
$21<V_0<21.4$ and $-0.2<(b-y)_0<0.5$ for the potential BHB
stars. The second region is given by the magnitude range above and below the first box, i.e. $15<V_0<21$ and $21.4<V_0<22$ with $-0.2<(b-y)_0<0.5$.

Figure \ref{plot28}a and b then show these stars both in the $c_1$ vs. $(b-y)$ and in the colour-magnitude diagram. The stars selected in the magnitude range of the HB are all narrowly spaced in magnitude and they all fall close to the BHB-RHB sequences. Hence we take all 10 of these stars to represent the HB of the Hercules dSph galaxy. We find a mean magnitude for the HB of $V_0=21.17\pm 0.05$ and $V=21.36\pm 0.05$.

	\subsection{A new distance determination to the Hercules dSph galaxy} \label{dist}
	
\begin{figure}
  \resizebox{\hsize}{!}{\includegraphics{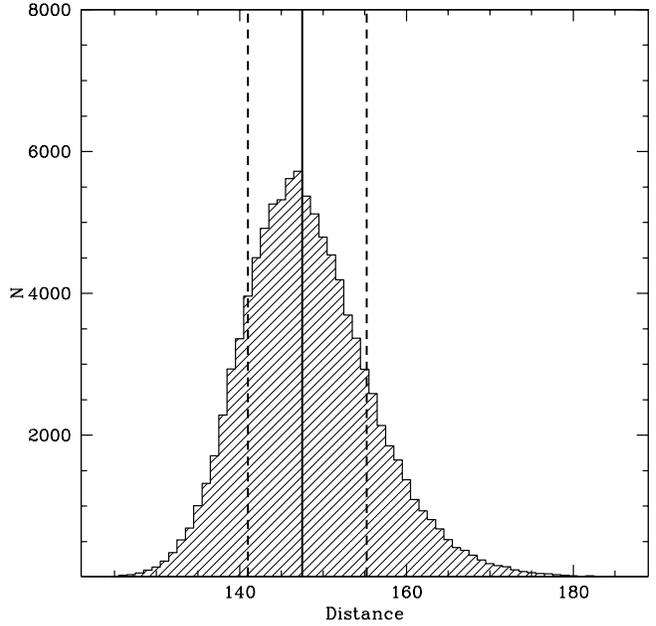}}
  \caption{Determination of the distance to the Hercules dSph galaxy, see Sect. \ref{dist}. Solid line indicates the median distance to the Hercules dSph galaxy. Dashed vertical lines indicate the upper and lower sextile.}
   \label{massmonte}
\end{figure}

We used the relation between the absolute magnitude of the HB and metallicity ($-2.35$ dex as derived in Sect. \ref{metal}), based on globular clusters, as defined in \cite{2000ApJ...533..215C} to determine the absolute magnitude for the HB, $M_{\rm HB}$, in the Hercules dSph galaxy. 
Using an apparent magnitude for the Hercules HB of $m_{\rm HB}=21.17\pm 0.05$ we solve for the distance, $d$
\begin{equation}
m_{\rm HB}-M_{\rm HB}=-5+5\cdot log(d)
\end{equation}

The errors in the distance determination were calculated using a Monte Carlo simulation. For each parameter with an associated error (apparent HB magnitude, metallicity and constants in the relation between the absolute magnitude and HB and metallicity), a new random apparent HB magnitude, metallicity and constants were calculated, and a new distance based on these parameters was found. This process was repeated 100000 times. In Fig. \ref{massmonte} we show the distribution of the Monte Carlo simulation. As the final errors for the distance, we adopt the lower and the upper sextile of the distribution of simulated distances; the lower sextile cuts off the lowest 16.67 per cent of the data and the upper sextile cuts off the highest 16.67 per cent of the data.
We find that Hercules is at a distance of $147^{+8}_{-7}$ kpc from us.

	\subsection{Variable stars in the Hercules dSph galaxy} \label{varia}

In Fig. \ref{plot28}b we see some scatter in the colour-magnitude diagram above the HB. In agreement with a Besan\c{c}on model, some of these objects are likely HB stars that belong to the Milky Way. However, the Besan\c{c}on model can not explain the overabundance of stars seen at $V\sim 20.4$ and $(b-y)_0\sim 0.35$.
We now investigate if these stars are variable stars that, due to variation in magnitude, position themselves above the HB. Since we do not have more than 3 exposures in the bright filters, $y$ and $b$, it is difficult to determine the variability of these stars as a function of time. 

We investigated the variance between the individual exposures in the $y,b,v$ and $u$ filter, but
given the small number of exposures we found no strong
indication of variability for these stars. \\

\citet{2006ApJ...649L..83S} report the finding of 15 RR Lyrae variable stars in the Bo\"otes dSph galaxy with periods from $\sim 0.3$ to $\sim 0.9$ days. Since Bo\"otes, with $M_V=-5.8$ \citep{2006ApJ...647L.111B}, is fainter than the Hercules dSph galaxy, finding a similar number of RR Lyrae stars in our sample is a plausible scenario. 

To further investigate the possible presence of variable stars we cross-correlated our photometry with the SDSS photometry in order to obtain $ugriz$ photometry for our stars. We then transformed the $ugriz$ magnitudes on to the Johnson-Cousins $UBVRI$ system using the colour transformations in \citet{2006A&A...460..339J}. \citet{2004AJ....127..861B} found 163 variable stars in the Draco dSph galaxy. Mean magnitudes in $V,I$ and $B$ and amplitudes in $V$ for these stars are available on-line. We de-reddened the Draco photometry using $E(B-V)=0.027$ \citep{2004AJ....127..861B} and constructed an ellipse encircling the distribution of the variable stars from \citet{2004AJ....127..861B} in the $V_0$ vs. $(B-V)_0$ colour-magnitude diagram (see Fig. \ref{plot28}c). 

Using isochrones by \citet{2008A&A...482..883M} we find that the HB for two metal-poor, old populations ($-2.3$ and $-2.0$ dex) coincide in $V$. We therefore normalized the positions of the Draco variable stars in $V_0$ so that the ellipse is aligned with the mean magnitude of our HB in the Hercules dSph galaxy, Fig. \ref{plot28}c. We also include the
variable stars in Draco from \citet{2004AJ....127..861B} that fall
outside the ellipse. We see that the positions of the variable stars
in the colour-magnitude diagram for the Draco dSph galaxy are similar
to the positions of our open triangles: 1) They have a similar colour
range 2) They are more likely to fall above the HB than below 3) The
spread in $V_0$ at $(B-V)_0\sim 0.5$ is similar for both data sets.

Our conclusion from this analysis is that the 8 stars, fainter than
$V_0=19$, likely are variable stars in the Hercules dSph galaxy. These
stars are listed in Table \ref{variable}.

\begin{table}
\caption{Possible variable stars in the Hercules dSph galaxy identified in Sect. \ref{varia}.}   
\label{variable}
\centering                             
\begin{tabular}{c c c c c }          
\hline\hline                        
ID & RA(2000) & DEC(2000) & $V_0$ & $(b-y)_0$   \\    
\hline                                  
11718 & 247.84445 & 12.60114  &  $ 20.44 \pm     0.02$  &    $0.44  \pm    0.04$      \\
22960 & 247.45022 & 12.59089  &  $ 20.19 \pm     0.02$   &   $0.33  \pm    0.04$      \\
33388 & 247.74828 & 12.96788  &  $ 20.38 \pm     0.04$    &  $0.13   \pm   0.06$      \\
41701 & 247.78819 & 12.78892  &  $ 20.74 \pm     0.03$   &   $0.36  \pm    0.04$      \\
41807 & 247.77753 & 12.76208  &  $ 20.34  \pm    0.02$   &   $0.14  \pm    0.04$      \\
42113 & 247.74934 & 12.76745  &  $ 20.41 \pm     0.02$   &   $0.25  \pm    0.04$      \\
42503 & 247.71592 & 12.77974  &  $ 21.78  \pm    0.04$   &   $0.05  \pm    0.06$      \\
43193 & 247.64148 & 12.79060  &  $ 20.54  \pm    0.03$   &   $0.39 \pm     0.04$      \\ 
\hline
\end{tabular}
\begin{list}{}{}
\item[] Column 1 lists the INT ID. 
Column 2 and 3 list their coordinates.
Column 4 and 5 list the Str\"omgren magnitude $V_0$ and colour $(b-y)_0$ and their associated errors, respectively.
\end{list}
\end{table}

	\section{Determining the systemic velocity of the Hercules dSph galaxy}  \label{vrmemb}

Using radial velocity measurements to separate the stars belonging to
a dSph galaxy from foreground stars has proven to be an efficient
method for membership determination. However, for the Hercules dSph
galaxy this method is complicated since this galaxy has a systemic
velocity that falls within the velocity distribution of the
foreground dwarf stars in the Milky Way. 

In Fig. \ref{besancon} we show the velocity distribution for our
observations together with a Besan\c{c}on model in the direction of
the Hercules dSph galaxy.  As can be seen, the velocity peak of the
Hercules dSph galaxy lies within the velocity distribution of the
Milky Way galaxy. A sample of member stars, identified as members
based only on radial velocity will therefore contain a non-negligible number of
foreground stars. However, adding knowledge about the 
evolutionary stage of the stars means that we can eliminate the foreground dwarf
stars (compare Sect. \ref{stsel}) and obtain a clean sample which can
be used to determine the systemic velocity and velocity dispersion for
the dSph galaxy.

In order to illustrate the importance of knowing the evolutionary
stage of the star we will first consider only the radial velocities as
a means to define a sample of stars belonging to the dSph galaxy.  
After that we will add knowledge about the evolutionary stage to clean
the sample further.

	\subsection{Using only radial velocities to select RGB members} \label{vronly}

We used the maximum likelihood method described in
\cite{2006AJ....131.2114W} to determine the mean heliocentric
velocity and internal velocity dispersion for stars in the direction of the Hercules dSph
galaxy.

The natural logarithm of the probability function defined in
\cite{2006AJ....131.2114W} was maximized

\begin{equation}
ln(p)=-\frac{1}{2}\sum_{i=1}^N ln(\sigma_i^2+\sigma_p^2)-\frac{1}{2}\sum_{i=1}^N\frac{(v_i-u)^2}{(\sigma_i^2+\sigma_p^2)}-\frac{N}{2}ln(2\pi)
\label{lnp}
\end{equation}

In each iteration we rejected stars deviating by more than $3\,
\sigma$ as they are likely not members.  As can be seen in
Fig. \ref{besancon}, the objects targeted with FLAMES span a broad
range of velocities. For the first pass through the maximum likelihood
iteration we selected stars with $10{\rm \, km\,
  s^{-1}}<V_{rad}<70{\rm \, km\, s^{-1}}$. The maximization converged
after 1 iteration.  This method gives us a mean velocity of $40.87 \pm
1.42 {\rm \, km\, s^{-1}}$ with a dispersion of $7.33 \pm 1.08 {\rm \,
  km\, s^{-1}} $.  Stars deviating by less than $3\, \sigma$ from the
velocity could be considered as possible Hercules dSph galaxy
members. We find 32 stars in this velocity range.
  
	\subsection{Weeding out foreground dwarf stars with the same velocity as the Hercules dSph galaxy} \label{weed}
	
\begin{figure}
\resizebox{\hsize}{!}{\includegraphics{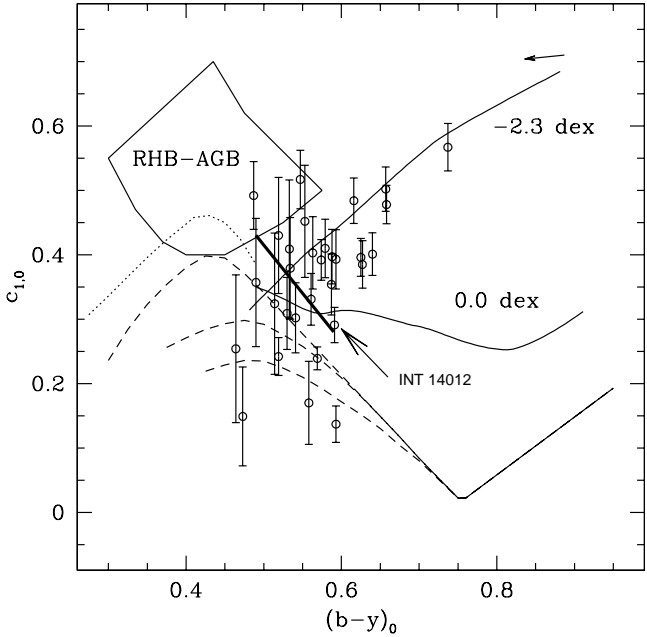}}
  \caption{$c_{1,0}$ vs. $(b-y)_0$ diagram for stars considered as members from the radial velocity measurement alone (see Sect. \ref{vronly}). The thin solid lines indicate the isochrones for RGB stars by \citet{2006ApJS..162..375V} with colour transformations by \citet{2004AJ....127.1227C}. The thin dashed lines indicate dwarf star sequences for different metallicities, ${\rm[Fe/H]}=0.45$, $-0.05$ and $-1$ top to bottom and the dotted line marks the upper envelope for dwarf stars (all lines from \'{A}rnad\'{o}ttir et al. in preparation). 
  The thick solid line is our lower limit for identification of RGB stars (see Sect. \ref{findrgb}). The arrow in the top right corner indicates the magnitude and direction of the
    de-reddening applied to the data (see Sect. \ref{red}).}
  \label{vrmembers}
\end{figure}

\begin{figure}
  \resizebox{\hsize}{!}{\includegraphics[angle=-90]{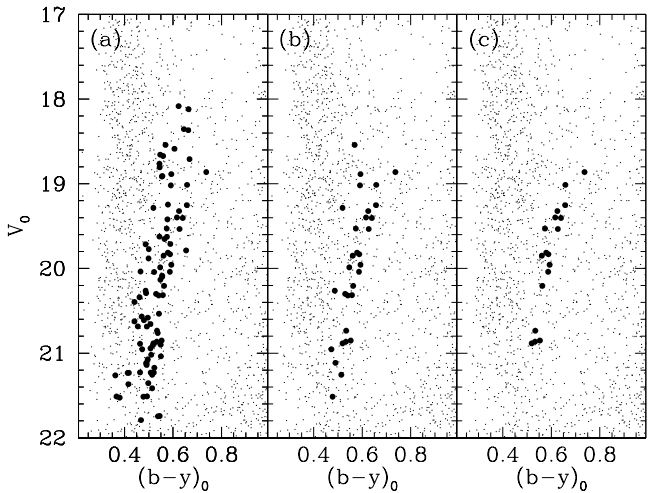}}
  \caption{Colour-magnitude diagrams. {\bf (a)} Objects targeted with FLAMES
    are shown as large $\bullet$. Small dots indicate stars from
    Fig. \ref{cmd4}a. {\bf (b)} Stars that have radial velocities within
    $3\sigma$ of the systemic velocity. {\bf (c)} Stars that are members
    based on the evolutionary stage and have measured velocities
    within $3\sigma$ of the systemic velocity (see Sect. \ref{weed}).}
  \label{plot38}
\end{figure}

One of the 32 stars found in Sect. \ref{vronly} (excluding the three stars with $sclass<0.5$), INT\,42568, lies on the edge of
  the WFC CCD \#4 and does not have any photometry available
and is hence excluded from the following discussions.
In Fig. \ref{vrmembers} we plot $c_{1,0}$ vs. $(b-y)_0$ for the 31  stars
considered as possible members based on the radial velocities.
We find the following

\begin{itemize}
\item Out of the 31 stars 10 fall on or below the dwarf sequences and are therefore excluded.
\item Of the remaining 21 stars, 18 are RGB
  members, 2 fall in the RHB-AGB region.

\item The last star falls below the RGB solar isochrone but is redder than
  the blue limit (compare Fig. \ref{plot3} and \ref{vrmembers}). This
  star is marked in Fig. \ref{vrmembers} (INT 14012).
  We consider this star as a member (but see discussion below).
\end{itemize}

Hence, we find about 30 per cent contamination by foreground stars in our sample. For the 19 RGB stars with the right evolutionary stage, we re-derive the mean velocity and dispersion using the maxiumum likelihood method.
We note that star INT 14012 was rejected during the iteration due to the $3 \sigma$ limit. This star is not included in the final sample of RGB stars.
We note that one of the remaining 18 stars, INT\,42170, has a velocity just outside the re-derived more narrow
$3\sigma$ limit. Since it falls short of this limit by only $1.32 {\rm \, km\,
s^{-1}} $ and the error in the velocity for the star is $4.77 {\rm \, km\,
s^{-1}} $, we keep it in the final sample. This star could also be a binary, which would
explain its deviating velocity. We found no stars that are likely members based on the Str\"omgren photometry, but non-members based on the radial velocity measurement.

We find a mean final systemic velocity of $45.20 \pm 1.09 {\rm \, km\, s^{-1}}$ with
a dispersion of $3.72 \pm 0.91 {\rm \, km\, s^{-1}} $. 
Our conclusion is that all stars, except INT 14012, with the right evolutionary stage fall within $3 \sigma$ of
the systemic velocity.

Figure \ref{plot38}a to c show the colour-magnitude diagrams for
objects targeted with FLAMES, stars identified as members from the
radial velocity measurement only and, finally, stars identified as
members from the radial velocity measurement and using photometry to
weed out foreground dwarf stars.

	\section{Final sample} \label{fs}

\begin{figure}
  \resizebox{\hsize}{!}{\includegraphics{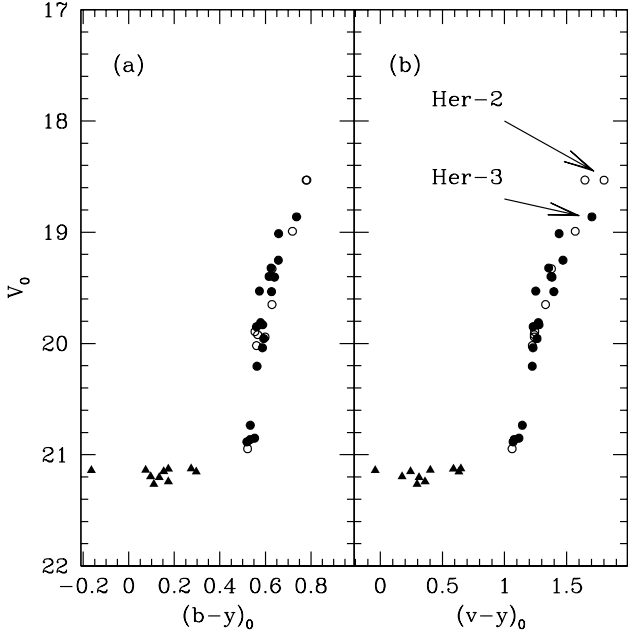}}
  \caption{Final colour-magnitude diagrams for the Hercules dSph galaxy. {\bf (a)} $V_0$ vs. $(b-y)_0$. {\bf (b)}
    $V_0$ vs. $(v-y)_0$. $\bullet$ are members based on the evolutionary stage and have measured velocities within $3\sigma$ of the systemic velocity (see Sect. \ref{weed}). $\circ$ are
    stars that do not have radial velocity measurements but are
    members according to their evolutionary stage (see Sect. \ref{stsel}). The filled triangles are stars identified as HB
    stars (see Sect. \ref{hb}). Her-2 and Her-3 are discussed in Sect. \ref{comp}.
    Note that there are two stars in {\bf (a)} at $V_0\sim 18.5$ that split in {\bf (b)}.}
  \label{plot22}
\end{figure}

Our final sample of RGB, AGB and HB Hercules member stars was defined as
follows: 1) first we select the RGB stars with the right evolutionary stage and a radial velocity within $\sim 3\sigma$ of the systemic velocity (compare Sect. \ref{weed} and
Fig. \ref{plot38}c). 2) To these stars we add stars without
spectroscopic measurements, which were selected as RGB, AGB, or
HB stars as determined from photometry (compare Sect. \ref{stsel}).
These stars are listed in Table \ref{photlist}.
Figure \ref{plot22}a and b show the colour-magnitude diagrams for the final
sample.

	\section{A comparison with previous velocity determinations} \label{vrcomp}

\begin{figure}
  \resizebox{\hsize}{!}{\includegraphics{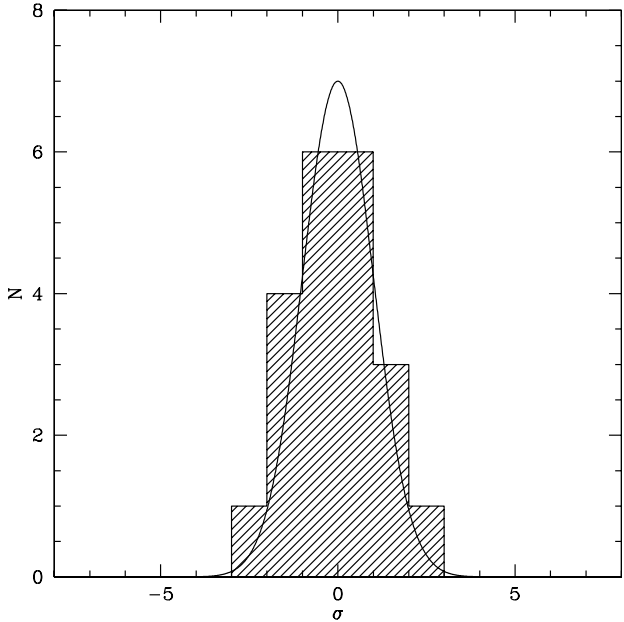}}
  \caption{Difference, in units of Gaussian $\sigma$ (see Sect. \ref{vrcomp}), between our measured stellar velocities and the velocities from \citet{2007ApJ...670..313S}. Note that this histogram includes both members and non-members.
  The solid line indicates a Gaussian with $\sigma=1$.}
   \label{simon_compare}
\end{figure}

\begin{figure}
  \resizebox{\hsize}{!}{\includegraphics{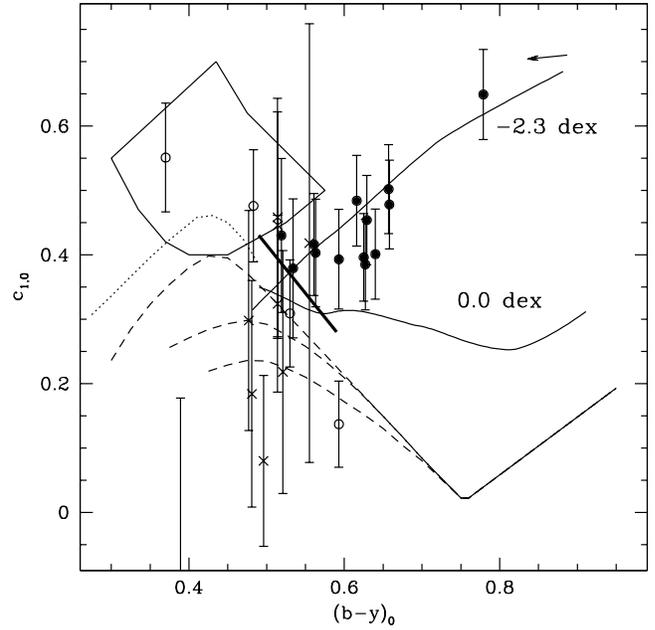}}
  \caption{
Str\"omgren $c_{1,0}$ vs. $(b-y)_0$ diagram for the stars
    from \citet{2007ApJ...670..313S}. $\bullet$ are stars
    considered as RGB members and $\circ$ are stars identified as
    non-RGB members based on our Str\"omgren photometry (see
    Sect. \ref{stsel}). $\times$ are stars that have $V_0>$21, hence
    we have not considered their evolutionary stage as they are too
    faint. The
    thin solid lines indicates the isochrones for RGB stars by
    \citet{2006ApJS..162..375V} with colour transformations 
    by \citet{2004AJ....127.1227C}. The thin dashed lines
    indicates dwarf star sequences for different metallicities,
    ${\rm[Fe/H]}=0.45$, $-0.05$ and $-1$ top to bottom and the dotted line
    marks the upper envelope for dwarf stars (all lines from \'{A}rnad\'{o}ttir et al. in preparation). The thick solid line is
    our lower limit for identification of RGB stars. The arrow in the top right corner indicates the magnitude and direction of the
    de-reddening applied to the data (see Sect. \ref{red}).}
   \label{simon1}
\end{figure}

\citet{2007ApJ...670..313S} have obtained radial velocities for
86 stars in the direction of the Hercules dSph galaxy. In order
to avoid including foreground dwarf stars they used measurements of
the strength of the Na\,{\sc i} lines at 818.3 and 819.5 nm to distinguish between dwarf and giant stars. As discussed
in \citet{1997ApJ...479..902S} the strength of these lines depend on the gravity
of the star. Hence, it enables a distinction between dwarf and giant
stars. Out of the 86 stars 29 were identified as members based on the measurements of
the strength of the Na\,{\sc i} lines.

There are 21 stars in common, including field dwarf stars, between our spectroscopic study and \citet{2007ApJ...670..313S}\footnote{J. Simon and M. Geha have kindly provided the relevant data so that we could do this analysis.}.
In Fig. \ref{simon_compare} we show the difference, in units of Gaussian $\sigma$, in stellar velocity for these stars \citep[see][for a similar study of the Draco dSph galaxy]{2002MNRAS.330..792K}. 
We note that the velocity measurements are in agreement, and that there are no obvious outliers, hence there is no
evidence for binaries in this sample. However, given the limited time sampling of the combined spectroscopy, and the fact that there are stars in our spectroscopic study for which
there are no velocity measurement in \citet{2007ApJ...670..313S}, we cannot rule out the presence of binaries in our study.
The presence of binaries could inflate the observed velocity dispersion by a small amount \citep[e.g.][]{1996AJ....111..750O}.

For the stars in common between our photometric study
and \citet{2007ApJ...670..313S} we show a $c_{\rm 1}$ vs. $(b-y)$ diagram
in Fig \ref{simon1}. From this comparison we find the following

\begin{itemize}

\item Out of the 29 stars considered as members in
\citet{2007ApJ...670..313S}, 2 fall in the RHB-AGB region and are
therefore excluded as RGB members.

\item Of the remaining 27 stars, 12 have $V_0>$ 21, hence we have not
considered their evolutionary stage as they are too faint. 4 out of the 12
stars with $V_0>$ 21 fall outside the limits of Fig. \ref{simon1}.

\item 2 of the remaining stars fall on or below the dwarf sequences
and are therefor excluded as members of the Hercules dSph galaxy.

\item Thus 13 stars remain that are considered as RGB members based
on Str\"omgren photometry.

\end{itemize}

For the 13 RGB members in common between us and \citet{2007ApJ...670..313S}, using the velocities from \citet{2007ApJ...670..313S}, we find a mean systemic velocity of $46.10 \pm 1.30 {\rm \, km\, s^{-1}}$ with a dispersion of $4.01 \pm 1.08 {\rm \, km\, s^{-1}}$. 

For 10 of the 13 RGB members we have determined radial velocities (see Sect. \ref{specmea}). Given the small offset in
velocity determination between our study and \citet{2007ApJ...670..313S}, see Fig. \ref{simon_compare}, it is possible to include the velocities of the 3 stars
for which we have not obtained a velocity measurement into our calculation of the mean velocity. We find a mean systemic velocity of $44.95 \pm 1.02 {\rm \, km\, s^{-1}}$ with a dispersion of $3.84 \pm 0.85 {\rm \, km\, s^{-1}}$ which is in agreement with our result from Sect. \ref{weed}.

There are 15 stars in our final sample of RGB members (based on the evolutionary
stage) that do not have radial velocities measured in \citet{2007ApJ...670..313S}.

Measuring the strength of the Na\,{\sc i} to exclude foreground dwarf stars is valid for $(V-I)>1$
\citep{2006ApJ...652.1188G,2008ApJ...689..958K}. This corresponds to $(b-y)>0.55$ 
and is confirmed by our comparison where the 16 stars that are bluer than
this limit are identified as either foreground dwarf stars or HB stars
by the Str\"omgren indexes. 

\begin{table*}
\caption{Summary of determination of systemic velocities, velocity dispersions and metallicities for the Hercules dSph galaxy.}
\begin{minipage}[t]{\columnwidth}
\label{table:6}
\centering
\renewcommand{\footnoterule}{}  
\begin{tabular}{l l c c c c c}
\hline\hline
 & Number of stars & $v_{sys}$  & $\sigma$  & $\langle {\rm [Fe/H]} \rangle $ & ${\rm [Fe/H]_{Median}}$\\ 
 &  & [${\rm \, km\, s^{-1}}$] & [${\rm \, km\, s^{-1}}$] & dex & dex \\
\hline
This study & 28 [RGB stars, $c_1$ sel.] & ... & ... &  $-2.35\pm 0.31$ & $-2.25^{+0.14}_{-0.31}$ \\ 
This study &  32 [Only $V_{rad}$ sel.] & $40.87\pm 1.42$ & $7.33 \pm 1.08$ & ... & ... \\
This study & 18 [RGB stars with $V_{rad}$] & $45.20 \pm 1.09$ & $3.72 \pm 0.91$ & $-2.34\pm 0.30$ & $-2.25^{+0.14}_{-0.20}$ \\
S$\&$G & 29 [Their calculation] & $45.0\footnote{Value from \citet{2007ApJ...670..313S}} \pm 1.1$ & $5.1^a \pm 0.9 $ & ... & ... \\
S$\&$G+$c_1$ & 13 [$c_1$ sel. sample] & $46.10 \pm 1.30$ & $4.01 \pm 1.08$ & ... & ... \\
Kirby et al. & 20 [Their sample] & ... & ... & $-2.58\footnote{Value from \citet{2008ApJ...685L..43K}} \pm 0.51$& ... \\
Kirby et al.+$c_1$ & 12 [$c_1$ sel. sample] & ... & ... & $-2.70 \pm 0.47$ & $-2.72^{+0.35}_{-0.40}$ \\
\hline
\end{tabular}
\end{minipage}
\begin{list}{}{}
\item[] Column 1 and 2 list the number of stars for each study and a short description of how the stars were selected. Note: S$\&$G indicates the sample of 29 stars from \citet{2007ApJ...670..313S}. Column 3 and 4 list the systemic velocities and velocity dispersions and their errors. Column 5 lists the mean metallicities. Column 6 lists the median metallicities. Errors for the median metallicity are the upper and lower quartile. For the calculations we use our metallicites based on Str\"omgren photometry, but for the \citet{2008ApJ...685L..43K} sample we use their metallicities.
\end{list}
\end{table*}

	\section{Metallicities}  \label{metal}

	\subsection{Metallicities based on Str\"omgren photometry} \label{fehm1}
	\subsubsection{Determination of metallicities}\label{detmet}
	
\begin{figure}
  \resizebox{\hsize}{!}{\includegraphics{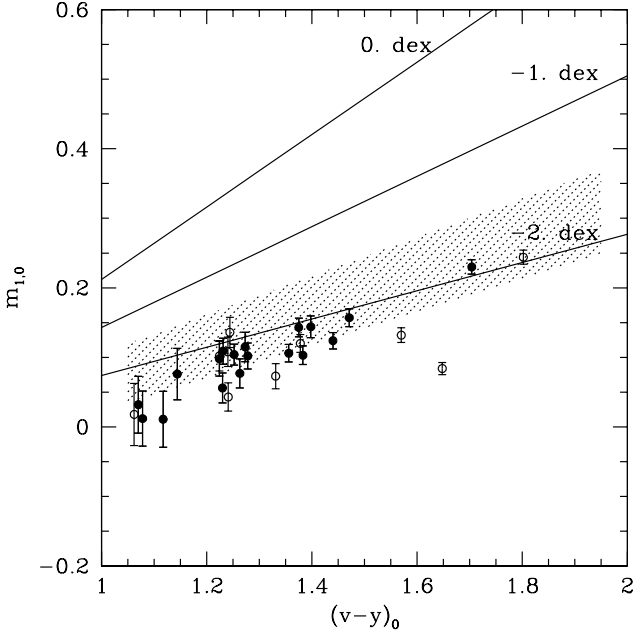}}
  \caption{$m_{1,0}$ vs. $(v-y)_0$ for the Hercules dSph
    galaxy. $\bullet$ are members based on the evolutionary stage as derived from photometry and they have radial velocities within $3\sigma$ of the systemic velocity for the Hercules dSph galaxy (see Sect. \ref{weed}). The $\circ$ are stars that do not
    have radial velocity measurements but are members according to our
    photometric criteria as developed in Sect. \ref{stsel}. The solid lines are iso-metallicity lines based on Eq. (\ref{FeHphotcalib2}) with metallicities as indicated. The shaded area indicates
    the distribution of RGB stars from \citet{2007A&A...465..357F} for the Draco dSph galaxy, adopted from their Fig. 19.}
  \label{plotm_1}
\end{figure}

The Str\"omgren filters have proven useful to estimate stellar
metallicities for RGB stars via the $m_1$ index
\citep[e.g.][]{1989A&A...211..199R},

\begin{equation} \label{m_1}
m_1=(v-b)-(b-y)
\end{equation}

A review of
the Str\"omgren metallicity calibrations for RGB stars, available at
the time, is provided by \citet{2007A&A...465..357F}. They adopt the
\cite{2000A&A...355..994H} calibration for their RGB stars in the Draco dSph galaxy. Since the \cite{2000A&A...355..994H} calibration is not valid
for ${\rm [Fe/H]}<-2.0$ dex and the newly found ultra-faint dSph galaxies such as
Hercules are metal-poor, we adopt the semi-empirical calibration by
\citet{2007ApJ...670..400C} onto the metallicity scale of \citet{1984ApJS...55...45Z} as this calibration is valid at least down to $-2.4$ dex. 

Figure \ref{plotm_1} shows the $m_{1,0}$ vs. $(v-y)_0$ plane for the
stars identified as RGB stars in the Hercules dSph galaxy.

Equation (\ref{FeHphotcalib2}), adopted from
\citet{2007ApJ...670..400C}, is used to convert $m_{1,0}$ and $(v-y)_0$
to ${\rm [Fe/H]_{Cal}}$ for the RGB stars.

\begin{equation} \label{FeHphotcalib2}
{\rm \left [Fe/H \right]_{Cal}} =\frac{(m_1+b_1 (v-y)+b_2 )}{(b_3 (v-y)+b_4 )}
\end{equation}

\noindent
where $b_1=-0.521 \pm 0.001$, $b_2=0.309$, $b_3=0.159 \pm 0.001$ and
$b_4=-0.09 \pm 0.002$ \citep[note that $b_2$ does not have an error estimate in][]{2007ApJ...670..400C}.

For comparison we re-calculated the stellar metallicities for the Draco RGB sample from \citet{2007A&A...465..357F}
using the calibration by \citet{2007ApJ...670..400C} (see Fig. \ref{plotm_1}). We found that the Draco dSph galaxy has a mean metallicity of $-2.0$ dex
instead of $-1.74$ dex as calculated 
by \citet{2007A&A...465..357F}.
A comparison between the metallicities derived using the older \cite{2000A&A...355..994H} calibration
and ${\rm [Fe/H]_{Cal}}$ shows an offset of $\sim 0.3$ dex, where ${\rm
  [Fe/H]_{Cal}}$ is more metal-poor. 

In Sect. \ref{red} we described how we correct the photometric
magnitudes for interstellar extinction using $E(B-V)=0.062$. As a test we re-calculated ${\rm [Fe/H]_{Cal}}$ using $E(B-V)=0.032$ and $E(B-V)=0.092$, which correspond to an uncertainty of $\pm 0.03$ mag.
We found that $E(B-V)=0.032$ increased and $E(B-V)=0.092$ decreased the metallicity by $\sim 0.1$ dex and $\sim 0.12$ dex, respectively.

	\subsubsection{Error estimates} \label{fehm1er}

\begin{figure*}
  \resizebox{\hsize}{!}{\includegraphics{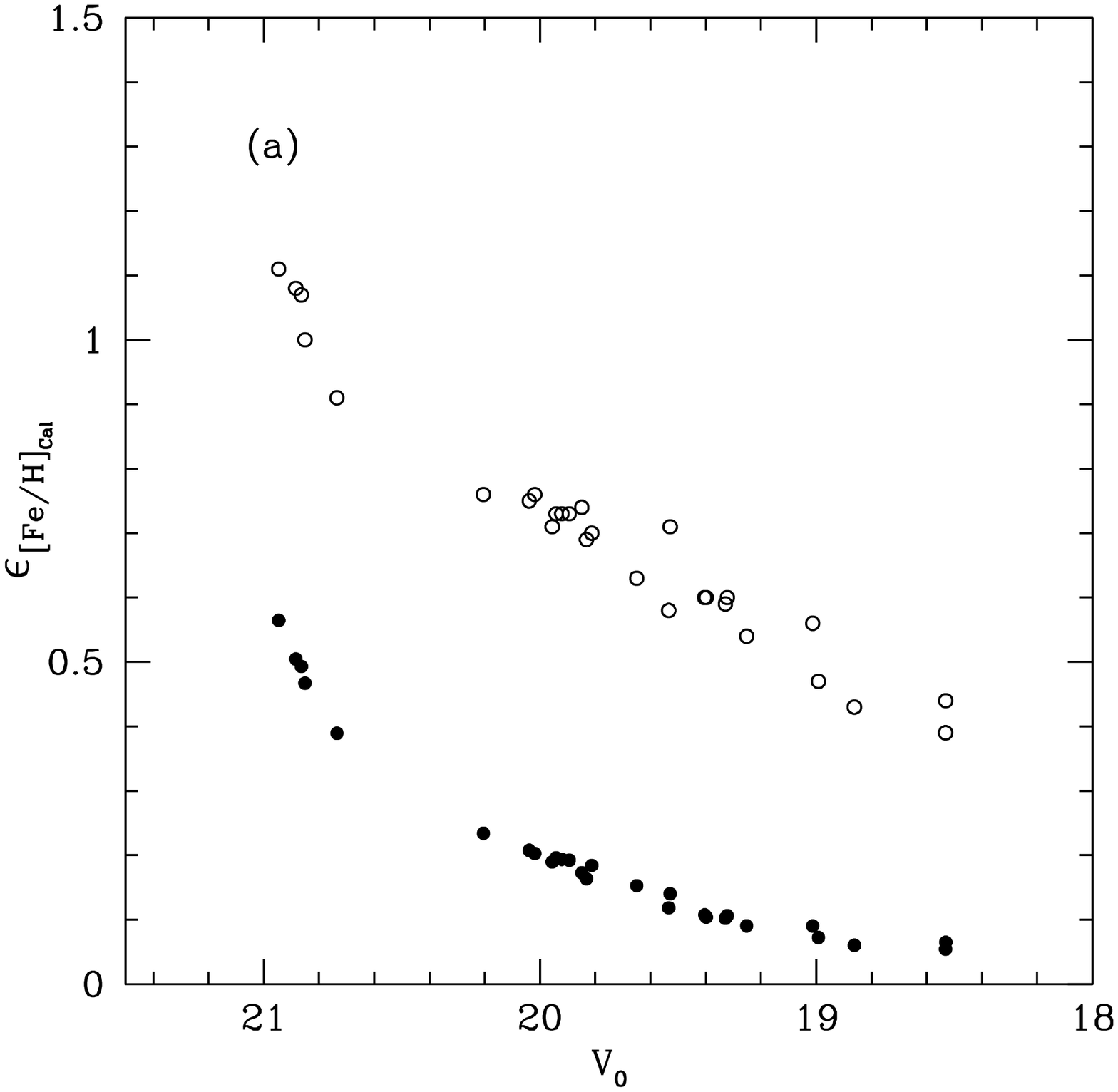}\includegraphics{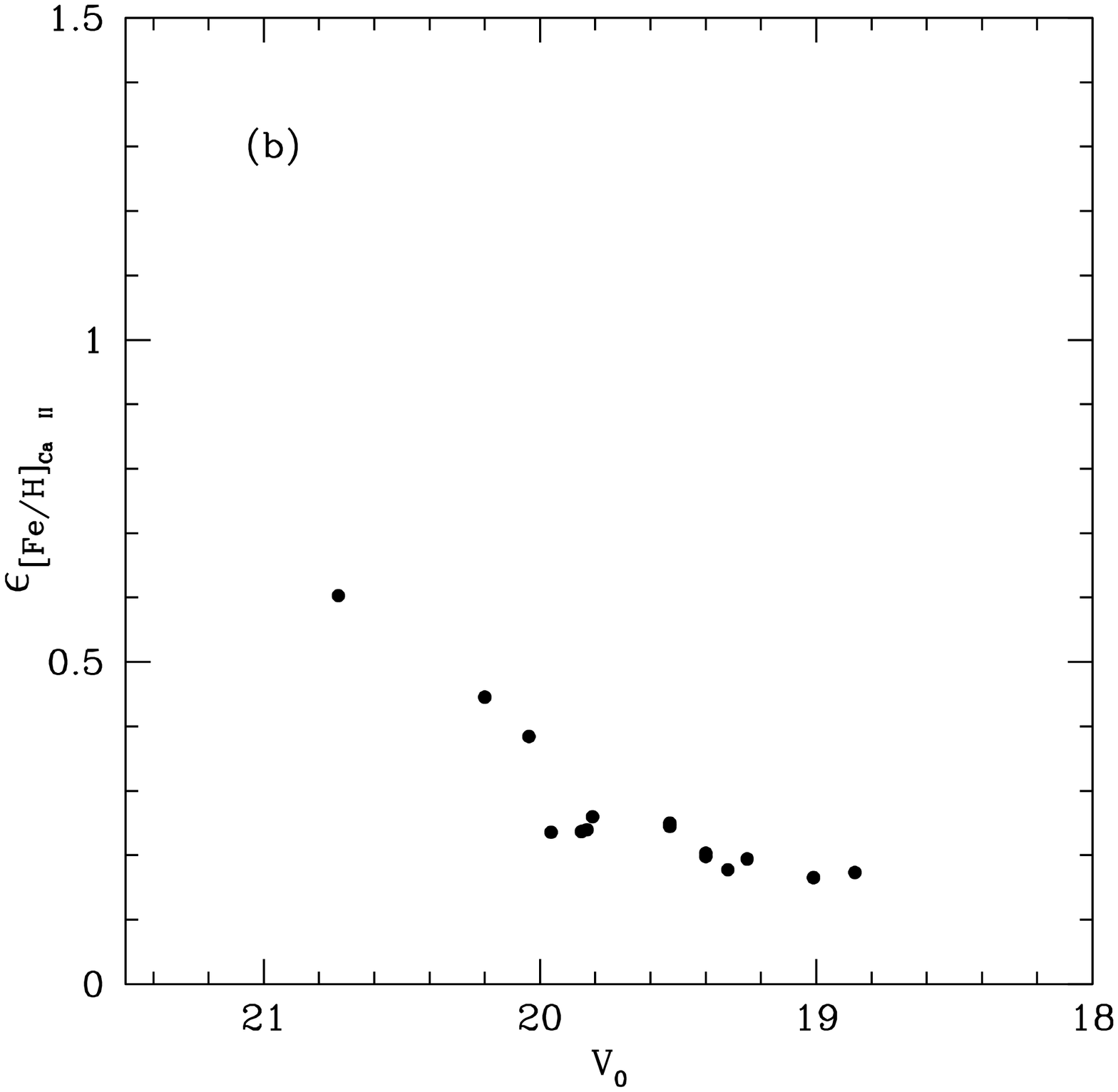}}
  \caption{{\bf (a)} Errors in the ${\rm[Fe/H]_{Cal}}$ for all stars identified
    as RGB members of the Hercules dSph galaxy in Sect. \ref{fehm1er}. 
    $\bullet$ indicates the errors if only {\tt merr} and the uncertainties in the coefficients of Eq. (\ref{FeHphotcalib2}) are included.
    $\circ$ indicates the errors if {\tt merr}, the uncertainties in the coefficients of Eq. (\ref{FeHphotcalib2}), and the
    uncertainties in zeropoints, extinction coefficients and colour terms are included.
     {\bf (b)} Errors in the ${\rm[Fe/H]_{Ca\, II}}$ as derived in Sect. \ref{fehcaiierror} for stars identified as RGB members .}
  \label{merr}
\end{figure*}

Following \citet{2007A&A...465..357F}, the errors in ${\rm [Fe/H]_{Cal}}$ were calculated using a Monte Carlo
simulation. 

The simplest version of this approach would be to calculate the errors in metallicities taking into account only the measurement errors from phot PHOT ({\tt merr}) and propagate those errors through Eq. (\ref{FeHphotcalib2}) \citep[this is the approach used in][]{2007A&A...465..357F}. 
A more involved approach would be to include also the uncertainties in zeropoints, extinction coefficients and colour terms. For comparison we have calculated the errors using both approaches.

In both cases we used the following approach to calculate the errors.
For every star identified as an RGB member of the Hercules
dSph galaxy, new random magnitudes ($v, b$ and $y$) were calculated
from Gaussian probability distributions with standard deviations
equal to that of the photometric error for each filter using a
Box-Muller transformation. $m_{1,0}$ and $(v-y)_0$ were re-calculated
from these new magnitudes and used to derive a new metallicity for the
star. This process was then repeated $10^5$ times for each star and since the
distributions of the new metallicities are not Gaussian around the
original [Fe/H], a standard deviation based on the sextiles (which is equivalent to $1\sigma$ in the case of a Gaussian distribution) was
calculated for each star from the distribution of simulated [Fe/H]. This
sextile standard deviation is our error in ${\rm [Fe/H]_{Cal}}$ (compare \citet{2007A&A...465..357F}).

In Fig. \ref{merr}a we show the errors in ${\rm [Fe/H]_{Cal}}$ as a function of $V_0$.
Note that the five faintest stars have significantly larger errors. Also note that the errors are much larger when the uncertainties in zeropoints, extinction coefficients and colour terms are taken into account.
We find that it is the errors in the $b$ filter that is the largest contributor to the error in ${\rm [Fe/H]_{Cal}}$ when the uncertainties in zeropoints, extinction coefficients and colour terms are taken into account (compare Table \ref{table2}).
The errors in the $b$ filter account for almost 50 per cent of the total error.

	\subsection{Metallicities based on Ca {\sc ii} IR triplet lines}
	\subsubsection{Determination of metallicities} \label{fehcaii}
	
  \begin{figure}
  \resizebox{\hsize}{!}{\includegraphics{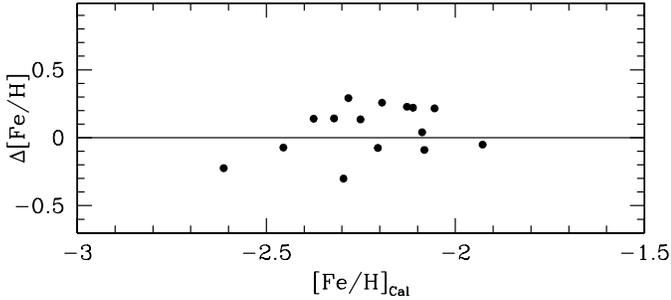}}
  \caption{A comparison between metallicities derived from the $m_1$
    index and those derived from Ca {\sc ii} IR triplet line
    measurements. $\Delta {\rm [Fe/H]}= {\rm [Fe/H]_{Cal}-[Fe/H]_{Ca\,
        II}}$. The offset is 0.06 dex with a scatter of 0.18 dex.}
   \label{plot11b}
\end{figure}

For those RGB stars for which we have FLAMES observations we also 
determined metallicities (${\rm [Fe/H]_{Ca\, II}}$) from measurements of the equivalent width of the Ca {\sc ii} IR
triplet lines at $\lambda=849.8$, $854.2$ and $866.2$ nm. We follow
\citet{1997PASP..109..883R} in defining the line strength of the Ca {\sc ii}
IR triplet lines as the weighted sum of the $W$, with lower
weights for the weaker lines

\begin{equation} \label{feh1}
\sum{W}=0.5 \cdot W_{1}+W_{2}+0.6 \cdot W_{3}
\end{equation}
where $W_{1}$, $W_{2}$ and $W_{3}$ are the widths of the individual Ca
II IR triplet lines in the order of increasing wavelength.

As discussed e.g. in \citet{1997PASP..109..907R}, the strength of the Ca {\sc ii}
IR triplet lines depend not only on metallicity but also on the surface
gravity and effective temperature of the star. It is possible to
remove the effect of gravity and temperature to first order by taking
into account the position of the star on the RGB. This is done by
defining the reduced $W$ as

\begin{equation} \label{feh2}
W'=\sum{W}+0.64 (\pm 0.02) (V-V_{HB})
\end{equation}
where $(V-V_{HB})$ is the difference between the $V$ magnitude of the star and the $V$ magnitude of the horizontal branch ($V_{HB}$).
The final ${\rm [Fe/H]_{Ca\, II}}$ were calculated using the calibration by \citet{1997PASP..109..907R} onto the metallicity scale of \citet{1997A&AS..121...95C}. The \citet{1997PASP..109..907R} calibration reads as follows
\begin{equation} \label{feh3}
{\rm \left [Fe/H \right]_{Ca \, II}} =-2.66(\pm 0.08)+0.42(\pm 0.02) W'
\end{equation}

Of the 18 RGB stars considered as members based on the evolutionary stage and which have measured velocities within $3\sigma$ of the systemic velocity, 3 have too low S/N for measurements of the $W$ and are therefore excluded from this metallicity determination. 

Figure \ref{plot11b} shows a comparison between our two metallicity estimates. We find that ${\rm [Fe/H]_{Cal}}$ is on
average 0.06 dex larger than ${\rm [Fe/H]_{Ca\, II}}$ with a scatter
of 0.18 dex. This scatter is consistent with the typical measurement uncertainty of ${\rm [Fe/H]_{Ca\, II}}$ (see below). In conclusion, the agreement between the photometric and spectroscopic
metallicities is very good. Since the calibration by \citet{2007ApJ...670..400C} is valid to at least [Fe/H]=$-2.4$, we conclude that
the extrapolation for the ${\rm [Fe/H]_{Ca\, II}}$ to ${\rm
  [Fe/H]}\sim -2.4$ is valid \citep[compare][]{2008MNRAS.383..183B}.
We note that the abundance scale of \citet{1984ApJS...55...45Z}, as used by  \citet{2007ApJ...670..400C}, in general has $\sim 0.2$ dex lower metallicities at around $-2.0$ dex than the abundance
 scale of \citet{1997A&AS..121...95C} as used by \citet{1997PASP..109..907R}.

	\subsubsection{Error estimates} \label{fehcaiierror}

Following \citet{2008MNRAS.383..183B}, we define the error due to random noise in the measurement of the $W$ as
\begin{equation} \label{ewerror}
\Delta W=\frac{\sqrt{1.5 \cdot {\rm FWHM}}}{{\rm S/N}}
\end{equation}
where FWHM is the Gaussian full-width-half maximum. The S/N varies from $\sim 24$ for the brightest star to $\sim 5$ for the faintest star.
The errors in ${\rm [Fe/H]_{Ca\, II}}$ were calculated following a Monte Carlo simulation procedure similar to the one used in Sect. \ref{fehm1er}. The process was repeated $100 \, 000$ times. As the final error for ${\rm [Fe/H]_{Ca\, II}}$, we adopt the standard deviation calculated from the distribution of simulated ${\rm [Fe/H]_{Ca\, II}}$.
In Fig. \ref{merr}b we show the errors in  ${\rm [Fe/H]_{Ca\, II}}$ as a function of $V_0$.

	\subsection{A comparison with metallicities determined in other studies} \label{comp}

\begin{figure}
  \resizebox{\hsize}{!}{\includegraphics{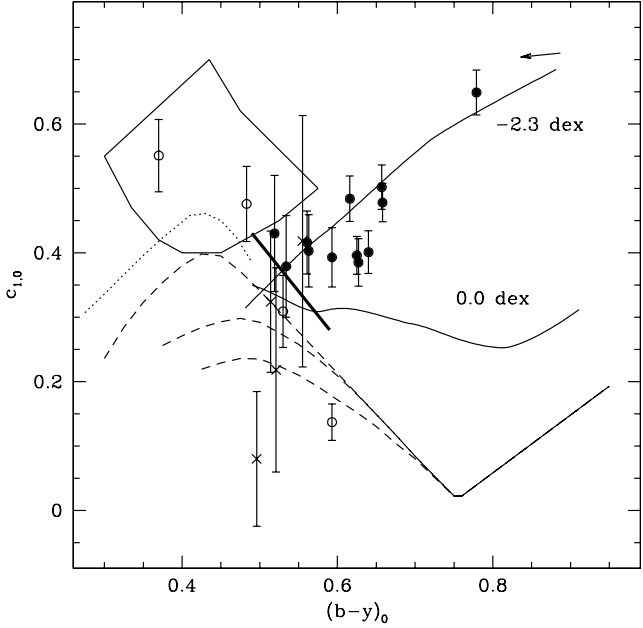}}
  \caption{ 
Str\"omgren $c_{1,0}$ vs. $(b-y)_0$ diagram for the stars
    from \citet{2008ApJ...685L..43K}. $\bullet$ are stars
    considered as RGB members and $\circ$ are stars identified as
    non-RGB members based on our Str\"omgren photometry (see
    Sect. \ref{stsel}). $\times$ are stars that have $V_0>$21, hence
    we have not considered their evolutionary stage as they are too
    faint for our membership determination based on Str\"omgren
    photometry. The
    thin solid lines indicates isochrones for RGB stars by
    \citet{2006ApJS..162..375V} with colour transformations
    by \citet{2004AJ....127.1227C}. The thin dashed lines
    indicates dwarf star sequences for different metallicities,
    ${\rm[Fe/H]}=0.45$, $-0.05$ and $-1$ top to bottom and the dotted line
    marks the upper envelope for dwarf stars (all lines from \'{A}rnad\'{o}ttir et al. in preparation). The thick solid line is
    our lower limit for the identification of RGB stars. The arrow in the top right corner indicates the magnitude and direction of the
    de-reddening applied to the data (see Sect. \ref{red}).}
   \label{kir2}
\end{figure}

\begin{figure}
  \resizebox{\hsize}{!}{\includegraphics{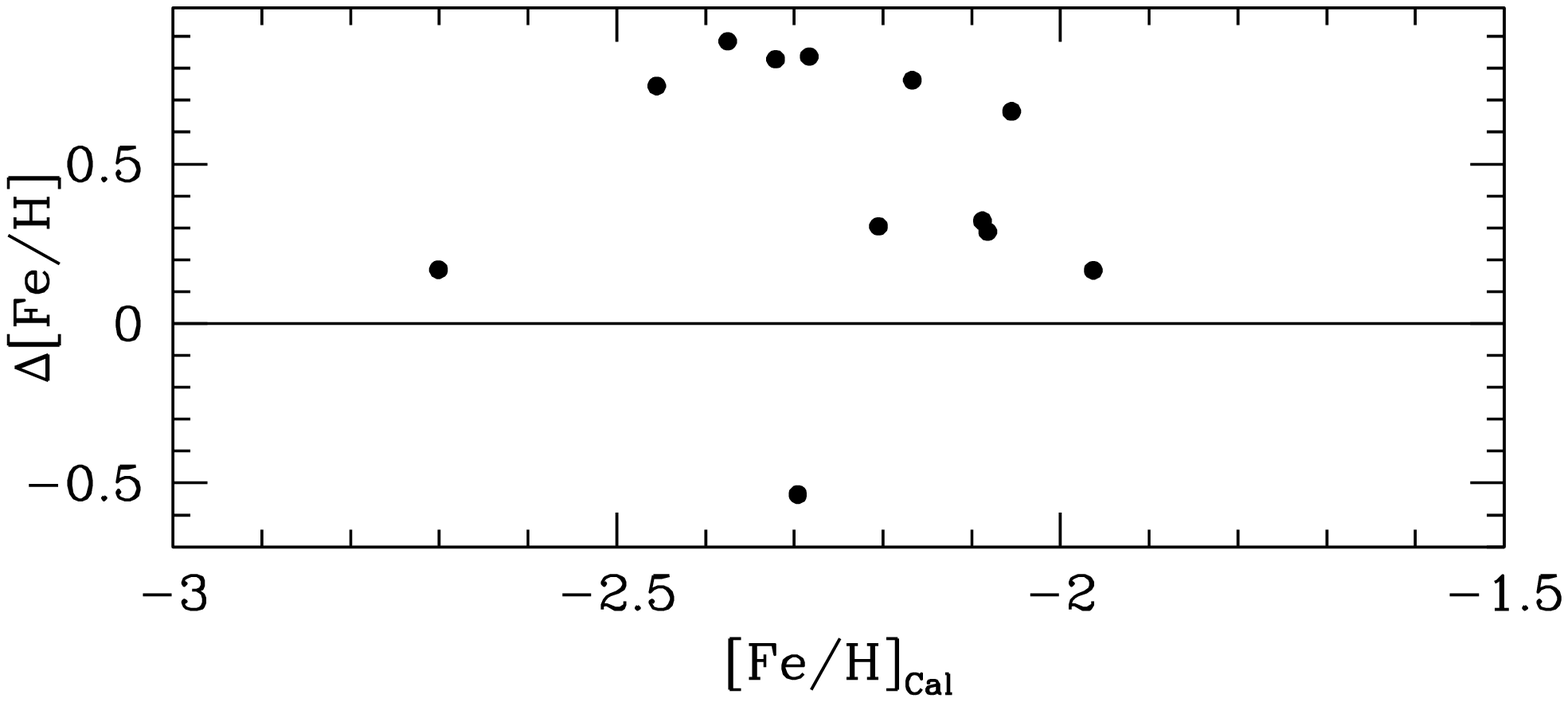}}
  \caption{A comparison between our metallicities derived from $m_1$
    and metallicities from \citet{2008ApJ...685L..43K}.  $\Delta {\rm
      [Fe/H]}= {\rm [Fe/H]_{Cal}-[Fe/H]_{Kirby}}$.}
   \label{kir}
\end{figure}

\citet{2008ApJ...688L..13K} obtained high resolution spectroscopy of
two stars in the Hercules dSph galaxy, Her-2 and Her-3. These stars
correspond to our stars INT 42241 ($V_0=18.53$) and INT 41082 ($V_0=18.86$). 
The stars are marked in Fig. \ref{plot22}. 
\citet{2008ApJ...688L..13K} find ${\rm [Fe/H]=-2.02}$ and
$-2.04$ for Her-2 and Her-3, respectively.

In Sect. \ref{stsel} we identify both of these stars as RGB stars and members of the Hercules dSph
galaxy. We only
have a radial velocity for one of the stars, INT 41082 (Her-3). The
velocity of this star falls within $3\sigma$ of our final systemic
velocity (see Sect. \ref{weed}). For this star we find ${\rm
  [Fe/H]_{Cal}}=-1.93$ and ${\rm [Fe/H]_{Ca \, II}}=-1.88$. For INT
42241 (Her-2) we derive ${\rm [Fe/H]_{Cal}}=-1.96$.

Hence, there is a difference of 0.11 and 0.06 dex, respectively, when
high resolution spectroscopy Fe abundances and metallicities derived
from Str\"omgren photometry are compared. This must be regarded as
excellent agreement given the complexities in analysing spectra of
such cool, evolved giant stars and the general simplifications made when
using calibrations of photometric measurements to obtain estimates of
stellar metallicities. It should also be noted that an overestimate in the reddening
of a few hundredths would easily account for this difference (compare Sect. \ref{fehm1}).

\citet{2008ApJ...685L..43K} studied 20 stars in the direction of the
Hercules dSph galaxy.  Their metallicities are based
on a recently developed automated spectrum synthesis method that
takes the information in the whole spectrum into account \citep{2008ApJ...682.1217K}. 
The method was originally developed for 
globular clusters in the Milky Way and was then applied to ultra-faint dSph galaxies in \citet{2008ApJ...685L..43K}.
[Fe/H] from \citet{2008ApJ...685L..43K} will
henceforth be referred to as ${\rm [Fe/H]_{Kirby}}$.
We have cross-correlated our photometry with
their 20 stars and found the following (see Fig. \ref{kir2}) \footnote{E. Kirby has kindly provided the
  relevant data so that we could do this analysis.}
\begin{itemize}
\item Out of the 20 stars considered as members by
  \citet{2008ApJ...685L..43K}, 2 fall in the RHB-AGB region and are
  therefore excluded as RGB members.
\item Of the remaining 18 stars, 4 have $V_0>$21, hence we have not
  considered their evolutionary stage but all indications are that at least
  three of them are foreground dwarf stars, compare Fig. \ref{kir2}. 
\item 2 of the 14 remaining stars fall on or below the dwarf sequences
  and are therefore foreground dwarf stars.
\end{itemize}

For the remaining 12 stars we find a mean metallicity of ${\rm [Fe/H]_{Cal}}=-2.25 \pm 0.20$ dex (Sect. \ref{fehm1}) and
a mean metallicity based on the values from
\citet{2008ApJ...685L..43K} ${\rm
  [Fe/H]_{Kir}}=-2.70 \pm 0.47$ dex. Figure \ref{kir}
shows the difference between  ${\rm [Fe/H]_{Cal}}$ and ${\rm [Fe/H]_{Kirby}}$.  We find that the median offset is 0.67\,dex, where ${\rm
  [Fe/H]_{Cal}}$ is more metal-rich. We have investigated $\Delta {\rm
  [Fe/H]}$ as a function of both magnitude and radial velocity but
found no trend.

\citet{2007ApJ...670..313S} obtained metallicities from 
the Ca {\sc ii} IR triplet lines and found a mean
metallicity of [Fe/H]=$-2.27$ dex. Finally,
\citet{2007ApJ...668L..43C} report a metallicity of [Fe/H]=$-2.26$ dex
based on fitting an isochrone to a colour-magnitude diagram.

In summary, we have five different determinations of the mean metallicity for 
the Hercules dSph galaxy. These determinations range from about $-2.25$ dex 
\citep[][this study]{2007ApJ...670..313S,2007ApJ...668L..43C} to --2.7 dex \citep{2008ApJ...685L..43K}. All determinations based 
on measurements of the Ca {\sc ii} IR triplet lines and the $m_{\rm 1}$-index agree on a
mean metallicity of about $-2.3$ dex. These determinations also agree with the high-resolution analysis by \citet{2008ApJ...688L..13K}.
The determinations by \citet{2008ApJ...685L..43K} are $\sim 0.5$ dex more metal-poor (Table \ref{table:6}). This discrepancy is not-negligible. However, future high-resolution studies of additional Hercules stars (as in \citet{2008ApJ...688L..13K}, which essentially is in agreement with our method) will determine which methods tend to over- or underestimate the [Fe/H].
This has important implications for our understanding of the origin of the metal poor halo stars \citep[e.g.][]{2006ApJ...651L.121H,2009arXiv0905.0557K}.

		\subsection{Metallicity for the Hercules dSph galaxy}
We find a metallicity range of $-2.99<{\rm [Fe/H]_{Ca \, II}}<-1.88$ with a mean metallicity
of ${\rm \langle [Fe/H]_{Cal}\rangle}= -2.35$ dex with $\sigma=0.31$
dex, and a median [Fe/H] of $-2.25$ dex with an upper and lower quartile of $-2.11$ dex and $-2.56$ dex, respectively. If we exclude stars fainter than $V_0 =20.5$ (compare Fig. \ref{merr}a) we find a mean metallicity
of ${\rm \langle [Fe/H]_{Cal}\rangle}= -2.26$ dex with $\sigma=0.24$
dex. We do not detect any significant spatial metallicity gradient in our data.

	\section{Discussion}  \label{disc}

	\subsection{Spatial distribution}  \label{disc3}

\begin{figure*}
  \resizebox{\hsize}{!}{\includegraphics{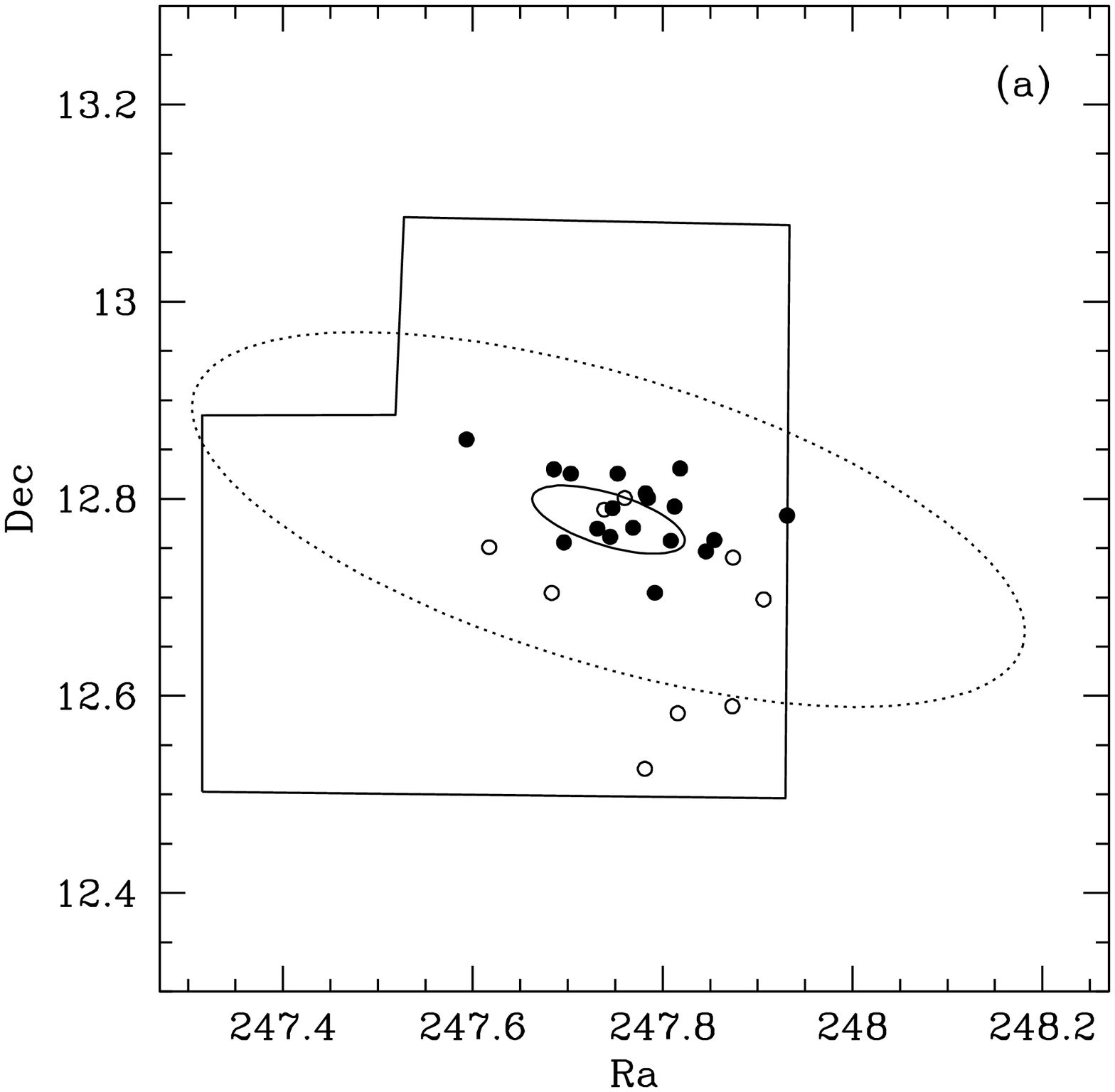}\includegraphics{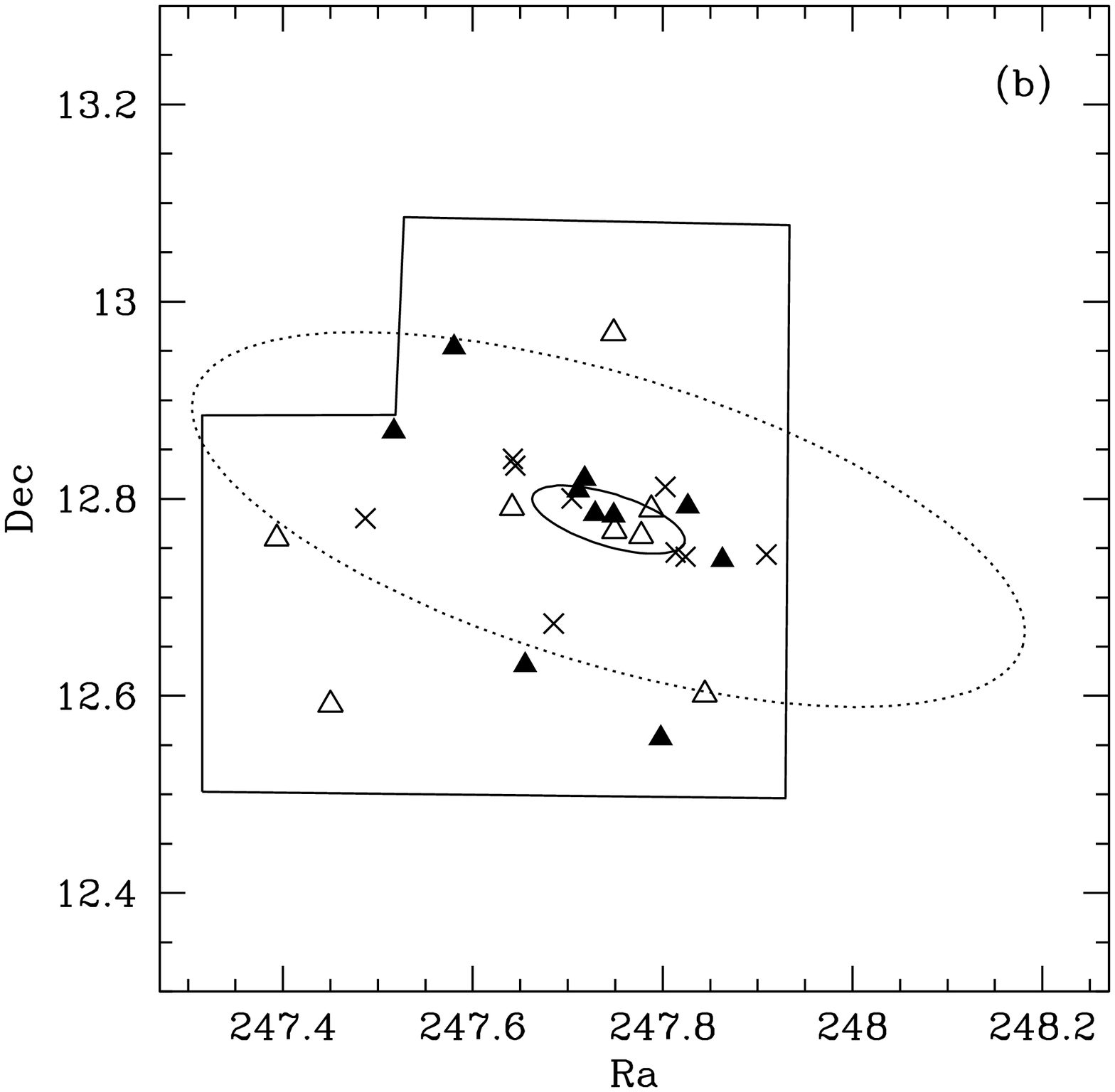}}
  \caption{Spatial distribution of the stars identified as members
    from both spectroscopy and photometry. {\bf (a)}$ \bullet$ are members
    based on the evolutionary stage as derived from photometry and
    they have radial velocities within $3\sigma$ of the systemic
    velocity for the Hercules dSph galaxy (see Sect. \ref{weed}). The
    $\circ$ are stars that do not have radial velocity measurements
    but are members according to our photometric criteria (see Sect. \ref{stsel}). {\bf (b)} Filled
    triangles are HB members. $\bigtriangleup$ are probable variable stars and $\times$ are stars identified as RHB-AGB based on the evolutionary stage as derived from photometry (see Sect. \ref{varia} and \ref{hb}, respectively).
    The solid ellipse
    represents the core radius and the dotted ellipse the King profile limiting radius
    of the Hercules dSph galaxy as determined by
    \citet{2007ApJ...668L..43C}. Solid lines outline the footprint of the WFC. Central coordinates for the galaxy are from \citet{2008ApJ...684.1075M}.}
  \label{plot4_spec}
\end{figure*}

The new, faint dSph galaxies are in general found to be quite
elongated \citep{2008ApJ...684.1075M}. The dSph galaxy in Hercules is no
exception, in fact it is one of the new galaxies with the largest
ellipticity ($e=0.68$). The large ellipticity of these objects might
be attributable to tidal distortions but could also be due to poor
sampling statistics. Each galaxy is only represented by a 
limited number of RGB stars in most of the cases \citep{2008ApJ...684.1075M}.
\citet{2008ApJ...684.1075M} and \citet{2008AN....329.1040U} conclude that it is entirely possible that 
the shape determinations for these the faintest of galaxies are
entirely dominated by shot-noise. This does not exclude tidal
disruption as an explanation for their, in general, very elongated 
shapes. 

With our Str\"omgren observations we have searched about 2/3 of the
area on the sky inside the King profile limiting radius, as defined by \citet{2007ApJ...668L..43C}, for members of the Hercules dSph galaxy.  Additionally, we have
searched about the same area on the sky outside the King profile limiting radius (see
Figs. \ref{camera} and \ref{plot4_spec}). Through this search we have found in total 28
stars that are RGB members of the Hercules dSph galaxy. These stars
are spatially confined to a fairly small, elongated area falling inside the core radius from
\citet{2007ApJ...668L..43C} (see Fig. \ref{plot4_spec}a). The RGB members show a slight
distortion such that there appear to be a few stars trailing the
ellipsoid in the direction of the Milky Way. These stars are few and
none have had their radial velocities measured.

The distribution of the HB, RHB, and potential variable stars are
shown in Fig. \ref{plot4_spec}b. These stars confirm the central concentration
and general shape of the Hercules dSph galaxy.
There are some HB, RHB and variable stars scattered outside
the ellipsoid defining the King profile limiting radius. It can not be excluded that
some of these stars are foreground contaminators. The presence of such
stars might distort the determination of the shape parameters for the
galaxy. Measurements of radial velocities for these stars would
confirm their membership. To our knowledge the HB of the Hercules
dSph galaxy has not yet been targeted for such observations.

In conclusion we find that for a well-defined sample of RGB stars
(free from contaminating foreground dwarf stars) the Hercules dSph
galaxy appears to still have a fairly elongated structure, confirming
previous studies. 

	\subsection{Measured velocity dispersion}  \label{disc4}

\begin{figure}
\resizebox{\hsize}{!}{\includegraphics{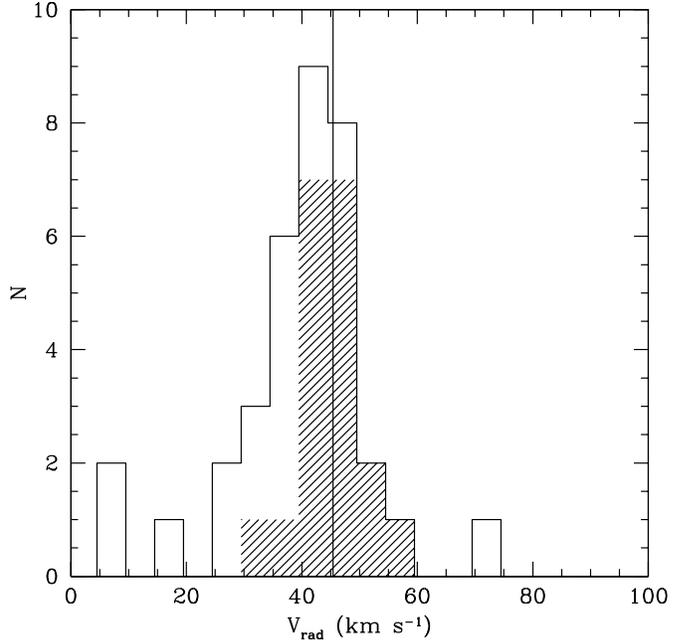}}
  \caption{Distributions of radial velocities. The solid histogram shows the distribution of velocities for objects in the direction of the Hercules dSph galaxy. The shaded histogram shows the distribution of radial velocities for stars identified as RGB members based on the evolutionary stage as derived from photometry. The solid-vertical line indicates the systemic velocity as derived from the RGB members.}
  \label{raddis}
\end{figure}

As discussed in Sect. \ref{vrmemb}, the velocity peak of the
Hercules dSph galaxy lies within the bulk of the velocity distribution of the
Milky Way galaxy, thus contaminating it. We have shown that relying on radial velocites
for membership determination yields a velocity dispersion of $7.33 \pm 1.08 {\rm \, km\, s^{-1}}$,
whilst excluding stars that are not members (i.e. they are foreground dwarf stars)
yields a velocity dispersion of $3.72 \pm 0.91 {\rm \, km\, s^{-1}}$.
\\

It is clear that the velocity dispersion is over-estimated when only considering
the radial velocity for membership identification. This is expected since the foreground contaminating 
stars in the direction of the Hercules dSph galaxy have a much broader velocity distribution than
that for the dSph galaxy (compare Fig. \ref{besancon}).

In Fig. \ref{raddis} we show the velocity distribution for our observations 
in the velocity range $0<V_{rad}<100 {\rm \, km\, s^{-1}}$, highlighting the objects 
that are members based on the evolutionary stage as derived in Sect. \ref{stsel}.
As can be seen, objects with a lower velocity than the systemic velocity 
of the Hercules dSph galaxy are more likely to be excluded as
foreground contaminating dwarf stars. 
Comparing with Fig. \ref{besancon} verifies that the 
majority of the velocities from a Besan\c{c}on model in the direction of the 
Hercules dSph galaxy have a velocity lower than the systemic velocity
for dSph galaxy. 

In conclusion we find that when deriving the velocity dispersion for dSph galaxies
it is important to have a well defined sample of member stars representing the 
velocity distribution of the dSph galaxy.

\subsection{Kinematic sub-structure in Hercules.}

\citet{2007ApJ...670..313S} detected possible evidence of kinematic
sub-structure in the Hercules dSph galaxy. They found nine stars clumped together
between 41 and 43 ${\rm \, km\, s^{-1}}$ in a sample of 30 stars
distributed between 30 and 60 ${\rm \, km\, s^{-1}}$. 

However, for our final sample of RGB stars in the Hercules dSph 
galaxy a one-sample Kolmogorov-Smirnov test yields a significance
level of 74$\%$ for the null hypothesis that the radial velocity distribution is
drawn from a Gaussian distribution. The conclusion is
therefore that we do not see any kinematic sub-structure in our sample.

\subsection{Metallicity} \label{fehdisc}

\begin{figure}
  \resizebox{\hsize}{!}{\includegraphics{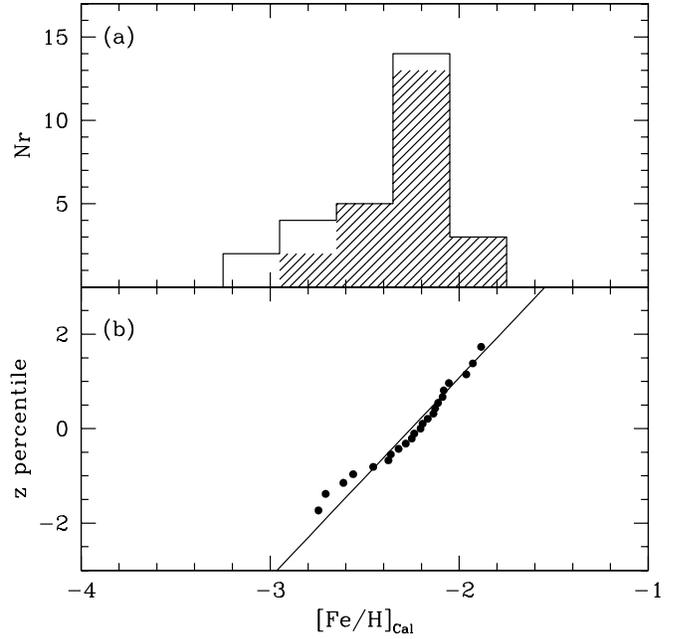}}
  \caption{{\bf (a)} Metallicity histogram for RGB stars in the Hercules dSph galaxy. The solid histogram shows the distribution for all stars in our final sample. The shaded histogram shows the distribution for stars in our final sample with $V_0 < 20.5$ (compare Fig. \ref{merr}a). {\bf (b)} Corresponding normal probability plot, for stars in our final sample with $V_0 < 20.5$, assuming a normal distribution. The solid line indicates a linear fit to the data with a mean metallicity of [Fe/H]=$-2.26$ dex and $\sigma=0.24$ dex.}
   \label{plot6}
\end{figure}

Figure \ref{plot6}a shows the metallicity distribution for the Hercules dSph galaxy based on the $m_1$-index for our final
sample of RGB members (see Sects. \ref{fs} and \ref{fehm1}). Figure \ref{plot6}b shows the corresponding
normal probability function. The data
points only deviate slightly from a linear fit which suggests that the metallicity distribution is
drawn from a single normal distribution. 
The slight deviation from the linear fit is noticeable at low metallicities where the cut-off is less sharp than at high metallicities.
These features is predicted by models and indicate the occurrence of strong galactic winds \citep{2007A&A...468..927L}.

In Sect. \ref{fehm1er} we calculated the errors in ${\rm [Fe/H]_{Cal}}$. When we take into account the uncertainties in zeropoints, extinction coefficients and colour terms we note that the errors are in general much larger than the $\sigma$ of the metallicity distribution (compare Fig. \ref{merr}a where we show the error in ${\rm [Fe/H]_{Cal}}$).
However, since the
uncertainties in zeropoints, extinction coefficients and colour terms propagate as a magnitude offset, equal for all stars, we re-calculate the errors in metallicities excluding the uncertainties in zeropoints, extinction coefficients and colour terms (see Fig. \ref{merr}a).
We find that these errors in metallicity, for stars brighter than $V_0 =20.5$,  are smaller than the metallicity spread of $0.24$ dex.
This enables us to study the star-to-star scatter in ${\rm [Fe/H]_{Cal}}$,
 making the profile of the distribution in Fig. \ref{plot6} significant.
 We conclude that there is an abundance spread in the metallicity distribution for the RGB members of up to 1 dex.

	\section{Summary}  \label{k}

We have, for the first time, presented a list of Hercules dSph galaxy
members based on an analysis of radial velocity, evolutionary stage
obtained from photometry, and stellar classification using SExtractor.
In detail we provide the following inventory of the Hercules
dSph galaxy
\begin{itemize}
\item 28 stars as RGB members based on their evolutionary stage, see
  Sect. \ref{stsel}. Of these, 19 have measured radial velocities (see Sect. \ref{weed}).
\item 9 stars as RHB-AGB members based on their evolutionary stage (see Sect. \ref{stsel}). Of these, 2 have radial velocities (see Sect. \ref{weed}).
 \item 10 stars as BHB members based on their evolutionary stage, see
  Sect. \ref{stsel}.
\item 8 stars that are possible variable stars based on their
  evolutionary stage, see Sect. \ref{varia}.
\end{itemize}

Our best determination of the systemic velocity is $45.20 \pm 1.09 {\rm \, km\, s^{-1}}$
with a dispersion of $3.72 \pm 0.91 {\rm \, km\, s^{-1}}$. We have
shown that membership based on radial velocity alone is not a good
method for the Hercules dSph galaxy, since it has a systemic velocity
that falls well within the velocity distribution of the foreground
dwarf stars belonging to the Milky Way.

Stellar metallicities have been determined using the
Str\"omgren $m_1$ index with a calibration that translates
$m_{1,0}$ to [Fe/H] for RGB stars. We found a mean metallicity of $-2.35$ dex. 
We also derived metallicities for our stars observed with 
FLAMES from measurements of the equivalent width of the Ca {\sc ii} IR triplet lines. 
The agreement between the two determinations 
was very good, with an offset of only 0.06 dex.

Finally, we have estimated the mean magnitude of the HB of the Hercules dSph
galaxy to $V_0=21.17\pm 0.05$ based on the 10 stars identified as BHB
members. This gives a distance of $147^{+8}_{-7}$ kpc.

\addtocounter{table}{1}

\begin{acknowledgements}
  We would like to thank Andreas Korn for putting us on the right
  track to understand the erroneous spectra created by the CCD-glow.
  D.A. thanks Simon Hodgkin at the Wide Field Survey unit at IoA,
  Cambridge, for help with the INT data reduction. D.A. thanks Lennart Lindegren
  at Lund Observatory for his help in problems of mathematical nature.
  S.F. is a Royal
  Swedish Academy of Sciences Research Fellow supported by a grant
  from the Knut and Alice Wallenberg Foundation.
  M.I.W. is supported by a Royal Society University
  Research Fellowship. \\
   Funding for the SDSS
  and SDSS-II has been provided by the Alfred P. Sloan Foundation, the
  Participating Institutions, the National Science Foundation, the
  U.S. Department of Energy, the National Aeronautics and Space
  Administration, the Japanese Monbukagakusho, the Max Planck Society,
  and the Higher Education Funding Council for England. The SDSS Web
  Site is http://www.sdss.org/.  The SDSS is managed by the
  Astrophysical Research Consortium for the Participating
  Institutions. The Participating Institutions are the American Museum
  of Natural History, Astrophysical Institute Potsdam, University of
  Basel, University of Cambridge, Case Western Reserve University,
  University of Chicago, Drexel University, Fermilab, the Institute
  for Advanced Study, the Japan Participation Group, Johns Hopkins
  University, the Joint Institute for Nuclear Astrophysics, the Kavli
  Institute for Particle Astrophysics and Cosmology, the Korean
  Scientist Group, the Chinese Academy of Sciences (LAMOST), Los
  Alamos National Laboratory, the Max-Planck-Institute for Astronomy
  (MPIA), the Max-Planck-Institute for Astrophysics (MPA), New Mexico
  State University, Ohio State University, University of Pittsburgh,
  University of Portsmouth, Princeton University, the United States
  Naval Observatory, and the University of Washington.

\end{acknowledgements}

\bibliographystyle{aa} 
\bibliography{12718}

\longtabL{9}{
\begin{landscape}
\begin{longtable}{c c c c c c c c c c c c c c c} 
\caption{\label{photlist} Hercules dSph galaxy members.}\\
\hline\hline
ID & RA(2000) & DEC(2000) & $V$ & $b$ & $v$ & $u$ & Chip &  & $[Fe/H]_{Cal}$ &  $V_{rad}  [ {\rm \, km\, s^{-1}}]$  \\ 
\hline
\endfirsthead
\caption{continued.}\\
\hline\hline
ID & RA(2000) & DEC(2000) & $V$ & $b$ & $v$ & $u$ & Chip &  & $[Fe/H]_{Cal}$ &  $V_{rad} [ {\rm \, km\, s^{-1}}]$ \\ 
\hline
\endhead
\hline
11239   & 247.87333   &  12.58958   &   19.92    $\pm$ 0.02     &    20.49    $\pm$ 0.04     &   21.16    $\pm$ 0.03     &   22.24    $\pm$ 0.04     &  1   &  RGB   &    -2.14    $\pm$ 0.73     &   ...   \\ 
12175   & 247.81591   &  12.58238   &   18.53    $\pm$ 0.02     &    19.31    $\pm$ 0.03     &   20.18    $\pm$ 0.02     &   21.59    $\pm$ 0.03     &  1   &  RGB   &    -2.71    $\pm$ 0.44     &   ...   \\ 
12729   & 247.78123   &  12.52606   &   19.65    $\pm$ 0.02     &    20.28    $\pm$ 0.03     &   20.98    $\pm$ 0.03     &   22.10    $\pm$ 0.04     &  1   &  RGB   &    -2.56    $\pm$ 0.63     &   ...   \\ 
40222   & 247.93108   &  12.78307   &   19.81    $\pm$ 0.02     &    20.39    $\pm$ 0.04     &   21.09    $\pm$ 0.03     &   22.19    $\pm$ 0.04     &  4   &  RGB   &    -2.13    $\pm$ 0.70     &    41.90    $\pm$ 3.65   \\ 
40457   & 247.90631   &  12.69785   &   20.95    $\pm$ 0.03     &    21.47    $\pm$ 0.04     &   22.01    $\pm$ 0.04     &   22.99    $\pm$ 0.12     &  4   &  RGB   &    -2.87    $\pm$ 1.11     &   ...   \\ 
40789   & 247.87404   &  12.74030   &   19.33    $\pm$ 0.02     &    19.96    $\pm$ 0.03     &   20.71    $\pm$ 0.03     &   21.91    $\pm$ 0.03     &  4   &  RGB   &    -2.24    $\pm$ 0.59     &   ...   \\ 
40993   & 247.85432   &  12.75811   &   19.53    $\pm$ 0.02     &    20.16    $\pm$ 0.03     &   20.93    $\pm$ 0.03     &   22.09    $\pm$ 0.04     &  4   &  RGB   &    -2.08    $\pm$ 0.58     &    40.22    $\pm$ 3.58   \\ 
41082   & 247.84564   &  12.74666   &   18.86    $\pm$ 0.02     &    19.60    $\pm$ 0.04     &   20.57    $\pm$ 0.02     &   22.10    $\pm$ 0.04     &  4   &  RGB   &    -1.93    $\pm$ 0.43     &    41.73    $\pm$ 0.67   \\ 
41371   & 247.81831   &  12.83070   &   20.04    $\pm$ 0.02     &    20.63    $\pm$ 0.04     &   21.27    $\pm$ 0.03     &   22.27    $\pm$ 0.04     &  4   &  RGB   &    -2.61    $\pm$ 0.75     &    46.11    $\pm$ 3.22   \\ 
41423   & 247.81247   &  12.79209   &   20.88    $\pm$ 0.03     &    21.40    $\pm$ 0.04     &   21.95    $\pm$ 0.04     &   22.93    $\pm$ 0.09     &  4   &  RGB   &    -2.70    $\pm$ 1.08     &    34.79    $\pm$ 5.06   \\ 
41460   & 247.80860   &  12.75741   &   19.40    $\pm$ 0.02     &    20.04    $\pm$ 0.04     &   20.79    $\pm$ 0.03     &   21.93    $\pm$ 0.03     &  4   &  RGB   &    -2.38    $\pm$ 0.60     &    48.20    $\pm$ 3.91   \\ 
41642   & 247.79179   &  12.70463   &   20.86    $\pm$ 0.03     &    21.40    $\pm$ 0.04     &   21.94    $\pm$ 0.04     &   22.90    $\pm$ 0.10     &  4   &  RGB   &    -2.96    $\pm$ 1.07     &    48.09    $\pm$ 8.30   \\ 
41737   & 247.78436   &  12.80075   &   19.94    $\pm$ 0.02     &    20.54    $\pm$ 0.04     &   21.18    $\pm$ 0.03     &   22.25    $\pm$ 0.04     &  4   &  RGB   &    -2.75    $\pm$ 0.73     &   ...   \\ 
41743   & 247.78386   &  12.80170   &   19.25    $\pm$ 0.02     &    19.91    $\pm$ 0.04     &   20.72    $\pm$ 0.02     &   22.04    $\pm$ 0.04     &  4   &  RGB   &    -2.09    $\pm$ 0.54     &    46.29    $\pm$ 0.95   \\ 
41758   & 247.78206   &  12.80545   &   20.20    $\pm$ 0.02     &    20.77    $\pm$ 0.04     &   21.43    $\pm$ 0.03     &   22.49    $\pm$ 0.05     &  4   &  RGB   &    -2.21    $\pm$ 0.76     &    43.18    $\pm$ 3.51   \\ 
41912   & 247.76877   &  12.77069   &   19.96    $\pm$ 0.02     &    20.55    $\pm$ 0.04     &   21.22    $\pm$ 0.03     &   22.28    $\pm$ 0.04     &  4   &  RGB   &    -2.46    $\pm$ 0.71     &    42.81    $\pm$ 2.78   \\ 
42008   & 247.76005   &  12.80071   &   20.02    $\pm$ 0.02     &    20.58    $\pm$ 0.04     &   21.24    $\pm$ 0.03     &   22.32    $\pm$ 0.04     &  4   &  RGB   &    -2.17    $\pm$ 0.76     &   ...   \\ 
42096   & 247.75261   &  12.82550   &   19.40    $\pm$ 0.02     &    20.01    $\pm$ 0.04     &   20.77    $\pm$ 0.03     &   22.02    $\pm$ 0.04     &  4   &  RGB   &    -2.06    $\pm$ 0.60     &    54.61    $\pm$ 1.62   \\ 
42149   & 247.74718   &  12.79045   &   19.01    $\pm$ 0.02     &    19.67    $\pm$ 0.04     &   20.45    $\pm$ 0.02     &   21.71    $\pm$ 0.04     &  4   &  RGB   &    -2.28    $\pm$ 0.56     &    44.48    $\pm$ 0.87   \\ 
42170   & 247.74480   &  12.76128   &   20.85    $\pm$ 0.03     &    21.40    $\pm$ 0.04     &   21.97    $\pm$ 0.04     &   22.98    $\pm$ 0.08     &  4   &  RGB   &    -2.99    $\pm$ 1.00     &    32.72    $\pm$ 4.77   \\ 
42241   & 247.73849   &  12.78898   &   18.53    $\pm$ 0.02     &    19.31    $\pm$ 0.04     &   20.33    $\pm$ 0.02     &   22.01    $\pm$ 0.04     &  4   &  RGB   &    -1.96    $\pm$ 0.39     &   ...   \\ 
42324   & 247.73111   &  12.76968   &   19.53    $\pm$ 0.02     &    20.10    $\pm$ 0.04     &   20.78    $\pm$ 0.03     &   21.85    $\pm$ 0.03     &  4   &  RGB   &    -2.19    $\pm$ 0.71     &    45.94    $\pm$ 2.18   \\ 
42637   & 247.70322   &  12.82538   &   20.73    $\pm$ 0.03     &    21.27    $\pm$ 0.04     &   21.88    $\pm$ 0.04     &   22.87    $\pm$ 0.07     &  4   &  RGB   &    -2.30    $\pm$ 0.91     &    49.41    $\pm$ 3.16   \\ 
42692   & 247.69607   &  12.75570   &   19.83    $\pm$ 0.02     &    20.42    $\pm$ 0.04     &   21.11    $\pm$ 0.03     &   22.20    $\pm$ 0.04     &  4   &  RGB   &    -2.25    $\pm$ 0.69     &    49.90    $\pm$ 1.45   \\ 
42795   & 247.68541   &  12.82996   &   19.32    $\pm$ 0.02     &    19.95    $\pm$ 0.03     &   20.68    $\pm$ 0.03     &   21.80    $\pm$ 0.03     &  4   &  RGB   &    -2.32    $\pm$ 0.60     &    42.90    $\pm$ 1.57   \\ 
42799   & 247.68313   &  12.70449   &   18.99    $\pm$ 0.02     &    19.71    $\pm$ 0.03     &   20.56    $\pm$ 0.03     &   21.86    $\pm$ 0.04     &  4   &  RGB   &    -2.36    $\pm$ 0.47     &   ...   \\ 
43428   & 247.61721   &  12.75078   &   19.89    $\pm$ 0.02     &    20.45    $\pm$ 0.04     &   21.14    $\pm$ 0.03     &   22.23    $\pm$ 0.04     &  4   &  RGB   &    -1.88    $\pm$ 0.73     &   ...   \\ 
43688   & 247.59341   &  12.86022   &   19.85    $\pm$ 0.02     &    20.41    $\pm$ 0.04     &   21.08    $\pm$ 0.03     &   22.08    $\pm$ 0.04     &  4   &  RGB   &    -2.11    $\pm$ 0.74     &    48.96    $\pm$ 2.14   \\ 
12453   & 247.79788   &  12.55701   &   21.20    $\pm$ 0.03     &    21.29    $\pm$ 0.04     &   21.37    $\pm$ 0.03     &   22.84    $\pm$ 0.06     &  1   &  HB    &   ...     &   ...   \\ 
14592   & 247.65505   &  12.63128   &   21.15    $\pm$ 0.03     &    21.30    $\pm$ 0.04     &   21.39    $\pm$ 0.03     &   22.86    $\pm$ 0.07     &  1   &  HB    &   ...     &   ...   \\ 
20190   & 247.51674   &  12.86830   &   21.14    $\pm$ 0.03     &    20.98    $\pm$ 0.04     &   21.10    $\pm$ 0.04     &   22.58    $\pm$ 0.07     &  2   &  HB    &   ...     &   ...   \\ 
35570   & 247.58018   &  12.95372   &   21.14    $\pm$ 0.03     &    21.21    $\pm$ 0.04     &   21.54    $\pm$ 0.03     &   22.84    $\pm$ 0.07     &  3   &  HB    &   ...     &   ...   \\ 
40911   & 247.86293   &  12.73797   &   21.15    $\pm$ 0.03     &    21.45    $\pm$ 0.04     &   21.79    $\pm$ 0.04     &   23.16    $\pm$ 0.10     &  4   &  HB    &   ...     &   ...   \\ 
41298   & 247.82626   &  12.79214   &   21.24    $\pm$ 0.03     &    21.41    $\pm$ 0.04     &   21.60    $\pm$ 0.03     &   22.94    $\pm$ 0.08     &  4   &  HB    &   ...     &   ...   \\ 
42134   & 247.74830   &  12.78317   &   21.20    $\pm$ 0.03     &    21.34    $\pm$ 0.04     &   21.52    $\pm$ 0.03     &   23.17    $\pm$ 0.12     &  4   &  HB    &   ...     &   ...   \\ 
42355   & 247.72872   &  12.78445   &   21.13    $\pm$ 0.04     &    21.30    $\pm$ 0.04     &   21.72    $\pm$ 0.04     &   22.92    $\pm$ 0.08     &  4   &  HB    &   ...     &   ...   \\ 
42484   & 247.71764   &  12.81998   &   21.27    $\pm$ 0.03     &    21.38    $\pm$ 0.04     &   21.56    $\pm$ 0.03     &   22.68    $\pm$ 0.06     &  4   &  HB    &   ...     &   ...   \\ 
42550   & 247.71108   &  12.80838   &   21.13    $\pm$ 0.03     &    21.40    $\pm$ 0.04     &   21.78    $\pm$ 0.03     &   22.87    $\pm$ 0.07     &  4   &  HB    &   ...     &   ...   \\ 
14151   & 247.68517   &  12.67333   &   20.99    $\pm$ 0.03     &    21.41    $\pm$ 0.04     &   21.77    $\pm$ 0.03     &   22.70    $\pm$ 0.05     &  1   &  RHB-AGB  &   ...     &   ...   \\ 
21089   & 247.48689   &  12.78015   &   20.41    $\pm$ 0.02     &    20.76    $\pm$ 0.04     &   21.13    $\pm$ 0.03     &   22.05    $\pm$ 0.04     &  2   &  RHB-AGB  &   ...     &   ...   \\ 
40435   & 247.90930   &  12.74345   &   20.26    $\pm$ 0.02     &    20.75    $\pm$ 0.04     &   21.30    $\pm$ 0.03     &   22.35    $\pm$ 0.05     &  4   &  RHB-AGB  &   ...     &    36.07    $\pm$ 3.27   \\ 
41310   & 247.82408   &  12.74113   &   19.99    $\pm$ 0.02     &    20.53    $\pm$ 0.04     &   21.13    $\pm$ 0.03     &   22.25    $\pm$ 0.04     &  4   &  RHB-AGB  &   ...     &    43.63    $\pm$ 5.72   \\ 
41401   & 247.81359   &  12.74562   &   20.22    $\pm$ 0.02     &    20.60    $\pm$ 0.04     &   21.03    $\pm$ 0.03     &   22.00    $\pm$ 0.04     &  4   &  RHB-AGB  &   ...     &   ...   \\ 
41532   & 247.80268   &  12.81217   &   20.45    $\pm$ 0.02     &    20.84    $\pm$ 0.04     &   21.24    $\pm$ 0.03     &   22.13    $\pm$ 0.04     &  4   &  RHB-AGB  &   ...     &   ...   \\ 
42621   & 247.70415   &  12.80033   &   20.61    $\pm$ 0.03     &    21.10    $\pm$ 0.04     &   21.56    $\pm$ 0.03     &   22.50    $\pm$ 0.05     &  4   &  RHB-AGB  &   ...     &   ...   \\ 
43167   & 247.64480   &  12.83356   &   19.92    $\pm$ 0.02     &    20.25    $\pm$ 0.04     &   20.54    $\pm$ 0.03     &   21.35    $\pm$ 0.03     &  4   &  RHB-AGB  &   ...     &   ...   \\ 
43194   & 247.64213   &  12.84056   &   20.72    $\pm$ 0.03     &    21.09    $\pm$ 0.04     &   21.48    $\pm$ 0.03     &   22.41    $\pm$ 0.05     &  4   &  RHB-AGB  &   ...     &   ...   \\ 
\end{longtable}
\begin{list}{}{}
\item[] Column 1 lists the ID. Column 2 and 3 list the coordinates. Column 4 to 7 list the Str\"omgren magnitudes $V$, $b$, $v$ and $u$. Column 8 lists on what CCD chip the star is located. Column 9 lists the evolutionary stage of the star. Column 10 lists the photometric metallicity as derived from the \cite{2007ApJ...670..400C} calibration. Column 11 list the radial velocity, if available.
\end{list}

\end{landscape}
}

\end{document}